\documentclass[11pt]{article}

\usepackage{etoolbox}
\newtoggle{biblatex}
\toggletrue{biblatex} 

\usepackage[T1]{fontenc}
\usepackage[utf8]{inputenc}
\usepackage{geometry}
\usepackage[small]{caption}
\usepackage{subcaption}
\usepackage{rotating}
\usepackage{array}
\usepackage[doublespacing]{setspace}
\usepackage{placeins} 

\usepackage{amsmath}
\usepackage{amsfonts}
\usepackage{fancyvrb}
\usepackage{graphicx}
\usepackage{mathpazo}

\usepackage{algorithmic}
\usepackage{algorithm}

\DeclareMathOperator\logit{logit}

\DeclareMathOperator\Chol{Chol }

\DeclareMathOperator\cov{cov}

 \newcommand{\Like}{f(y|\theta)}
 \newcommand{\Prior}{\pi(\theta)}
 \newcommand{\Post}{\pi(\theta|y)}
 \newcommand{\LikePrior}{f(y|\theta)\pi(\theta)}
 
 \newcommand{\Gtheta}{g(\theta)}
 
 \newcommand{\Phitheta}{\Phi(\theta|y)}
 \newcommand{\Ly}{\mathcal{L}(y)}
 \newcommand{\Dy}{\mathcal{D}(\theta,y)}

 \newcommand{\pkg}[1]{\textit{#1}}
\newcommand{\proglang}[1]{\textsf{#1}}
\newcommand{\func}[1]{\texttt{#1}}

\DeclareGraphicsExtensions{.png,.pdf}

\iftoggle{biblatex}{%

\usepackage[style=authoryear,%
			backend=bibtex,%
			maxbibnames=99,%
                        bibencoding=utf8,%
                        maxcitenames=1,
                        citetracker=true,
                        dashed=false,
                        maxalphanames=1,%
                        backref=false,%
			doi=false,%
			isbn=false,%
                        mergedate=basic,%
                        dateabbrev=true,%
                        natbib=true,%
                        uniquename=false,%
                        uniquelist=false,%
                        useprefix=true,%
                        firstinits=false
			]{biblatex}

 \addbibresource{./gds.bib}

\setlength{\bibitemsep}{1em}
\AtEveryCitekey{\ifciteseen{}{\defcounter{maxnames}{2}}}
\DeclareFieldFormat[article,incollection,unpublished]{title}{#1} 
\DeclareFieldFormat[thesis]{title}{\mkbibemph{#1}} 
\DeclareFieldFormat{pages}{#1} 
\renewbibmacro{in:}{%
  \ifentrytype{article}{}{ 
  \printtext{\bibstring{in}\intitlepunct}} 
}
\renewbibmacro*{volume+number+eid}{%
  \printfield{volume}%
  \printfield{number}%
  \setunit{\addcomma\addspace}%
  \printfield{eid}}

\DeclareFieldFormat[article]{volume}{\mkbibbold{#1}}
\DeclareFieldFormat[article]{number}{\mkbibparens{#1}}
\DeclareFieldFormat[article]{date}{#1}
\AtEveryBibitem{\clearfield{day}}

\renewbibmacro*{issue+date}{
  \printtext{%
    \printfield{issue}\addspace%
    \newunit%
}
  \newunit}

\renewbibmacro*{publisher+location+date}{
  \printlist{location}%
  \iflistundef{publisher}
    {\setunit*{\addcomma\space}}
    {\setunit*{\addcolon\space}}%
  \printlist{publisher}%
  \setunit*{\addcomma\space}%
  \newunit}

\renewcommand{\bibitemsep}{1ex}

}

\nottoggle{biblatex}{%
\usepackage{natbib}
\bibpunct[, ]{(}{)}{;}{a}{}{,}
\bibliographystyle{ormsv080}
\newcommand{\printbibliography}{\bibliography{gds}}
}

\title{Scalable Rejection Sampling for Bayesian Hierarchical Models}
\author{Michael Braun\\
Cox School of Business\\
Southern Methodist University\\
Dallas, TX 75275\\
braunm@smu.edu
\and
Paul Damien\\
McCombs School of Business\\
University of Texas at Austin\\
Austin, TX 78712\\
paul.damien@mccombs.utexas.edu
}
\date{November 2, 2014}

\begin{document}

\begin{singlespace}
\maketitle
\begin{center}\bfseries
Forthcoming in \emph{Marketing Science}\\
Special Issue on Big Data:  Integrating Marketing, Statistics and
Computer Science
\end{center}
\vspace{.4in}
\begin{abstract}
Bayesian hierarchical modeling is a popular approach to capturing
unobserved heterogeneity across individual units.  However, standard estimation methods such as Markov
chain Monte Carlo (MCMC) can be impracticable for modeling outcomes
from a large number of units.  We develop a new method to sample from posterior
distributions of Bayesian models, without using MCMC.  Samples are
independent, so they can be collected in parallel, and we do not need to be concerned with issues like chain
convergence and autocorrelation.  The algorithm is scalable under
the weak assumption that individual units are conditionally
independent, making it applicable for large datasets. It can also be
used to compute marginal likelihoods.
\end{abstract}
\end{singlespace}

\thispagestyle{empty}
\newpage
\setcounter{page}{1}\section{Introduction}

In 1970, John D. C. Little famously wrote:  ``The big problem with
management science models is that managers practically never use
them.  There have been a few applications, of course, but the practice
is a pallid picture of the promise'' \citep{Little1970}.  The same may be true 
 today about Bayesian estimation of hierarchical probability
models. The impact Bayesian methods have had on
academic research across multiple disciplines in the managerial,
social and natural sciences is undeniable. Marketing, in particular, has benefited
from Bayesian methods because of their
natural suitability for capturing heterogeneity in customer types and
tastes \citep{RossiAllenby2003}.  But further diffusion of Bayesian methods is constrained by a
scalability problem.  As the size and complexity of data sources for
both research and commercial purposes grows, the impracticality of
simulation-based Bayesian methods for estimating
parameters of a general class of hierarchical models becomes
increasingly salient \citep{AllenbyBradlow2014}. 

The problem is not with the Bayesian approach itself,
but with the most familiar methods of simulating from the posterior
distributions of the parameters.  Without question, the resurgence of Bayesian ideas
is due to the popularity of Markov chain Monte Carlo (MCMC),
which was introduced to statistical researchers by
\citet{GelfandSmith1990} via the Gibbs sampler.  MCMC estimation involves
iteratively sampling from the marginal posterior distributions of
blocks of parameters.  Only after some unknown (and theoretically infinite)
number of iterations will the algorithm generate samples from the
correct distributions; earlier samples are discarded.  The Bayesian
computational literature has 
exploded with numerous methods for generating valid and efficient MCMC
algorithms.  It would be difficult to list all of them here, so we refer
the reader to \citet{GCSR2004,ChenShao2000,RossiAllenbyMcCulloch2005} and
\citet{BrooksGelman2010}, and the hundreds of references
therein. 

Despite the justifiable success MCMC has enjoyed, there remains the
question of whether a particular chain has run long enough that we can start collecting samples for
estimation (or, colloquially, whether the chain has ``converged'' to
the target distribution).   This is a particular problem for hierarchical models in which each
heterogeneous unit is characterized by its own set of parameters.  For
example, each household in a customer dataset might have its own
preferences for product attributes.  Both the number of
parameters and the cycle time for each MCMC iteration grow with the
size of the dataset.  Also, if the data
represent outcomes of multiple interdependent processes (such as the
timing and magnitude of purchases), both the posterior
parameters and successive MCMC samples tend to be correlated,
requiring a larger, yet unknown, number of
iterations. We believe that the most important reason Bayesian methods have not been embraced ``in the field''
nearly as much as classical approaches is that they are difficult
and expensive to implement routinely using MCMC, even with
semi-automated software procedures.  Practitioners simply
do not have an 
academician’s luxury of letting an MCMC chain run for days or weeks 
with no guarantee that the chain has converged to
produce “correct” answers at the end of the process.  

With recent developments in multiple core processing and distributed
computing systems, it is reasonable to look to parallel computing
technology as a solution to the convergence problem.  However, each MCMC
cycle depends on the outcome of the previous one, so we cannot collect posterior samples in parallel by
allocating the work across distributed processing cores.   Using
parallel processors to generate one draw from a target distribution,
or running several MCMC chains in parallel, is not the same as generating all the required independent
samples in parallel. Hence, extant parallel MCMC methods are also subject to
the same pesky question of convergence; indeed, now one has to ensure
that all of the parallel chains have converged. On the other hand, non-MCMC methods like rejection sampling have the
advantage of being able to generate samples from the correct target posterior in
parallel, but these methods are beset with their own set of
implementation issues.  For instance, the inability to find efficient ``envelope'' distributions
renders standard rejection sampling almost impractical for all but the smallest problems. 

In this paper, we propose a solution to sample from Bayesian
parametric, hierarchical models that is inspired by two pre-MCMC
approaches:  rejection sampling, and sampling from a multivariate
normal (MVN) approximation around the
posterior mode. Our contribution is an algorithm that recasts
traditional rejection sampling in a way that circumvents the
difficulties associated with these two approaches. The algorithm
requires that 
 one be able to compute the unnormalized
log posterior of the parameters (or a good approximation of it); that the posterior
distribution is bounded from above over the parameter space; and that available computing resources
can locate any local maxima of the log posterior.  There is no need to derive
conditional posterior distributions (as with blockwise Gibbs
sampling), and there are no conjugacy requirements.  

We present the details of our method in Section
\ref{sec:GDS}, and in Section \ref{sec:empirical}, we share some examples
that demonstrate the method's effectiveness. In broad strokes, the
method involves scaling an MVN distribution around the mode, and using that
distribution as the source of proposal draws for the modified rejection
algorithm.  At first glance, one might think that finding the
posterior mode, and sampling from an MVN, are themselves intractable tasks
in large dimensions. After all, the Hessian of the log posterior
density, which grows quadratically with the number of parameters, is
an important determinant of the efficiency of both MVN sampling and
nonlinear optimization.  Fortunately, several
independent software development projects have spawned novel, freely
available numerical computation tools that, when used together, allow
our method to scale.  In Section
\ref{sec:scalability}, we explain how to manage this scalability
issue, and show that the complexity of our
method scales approximately linearly with the number of heterogeneous units.

Another complication of Bayesian methods is the estimation of the marginal
likelihood of the data.  The marginal likelihood is the probability
of observing the data under the proposed model, which can be used as a metric for model comparison. Except
in rare special cases, computing the marginal
likelihood involves numerically integrating over all of the prior
parameters; note that we consider hyperpriors to be part of the
data in this case.  In Section \ref{sec:MargLL}, we explain how to
estimate the marginal likelihood as a by-product of our method.

In Section \ref{sec:practical}, we discuss key implementation issues,
and identify some relevant software tools.  We also discuss 
 limitations of our approach.  We are not claiming
that our method should replace MCMC in all cases.  It may not
be practical for models with discrete parameters, a very large number
of modes, or for which computing the log posterior density itself is
difficult.  The method does not require that the model be
hierarchical, or that the conditional independence assumption holds,
but without those assumptions, it will not be as scalable.  Nevertheless, many models of the kind researchers encounter could be properly 
estimated using our method, at least relative to the effort involved in
using MCMC. Like MCMC and other non-MCMC methods, our method is another
useful algorithm in the researcher's and practitioner's toolkits.

\section{Method details}\label{sec:GDS}
\subsection{Theoretical basis}
The goal is to sample a parameter vector $\theta$ from a posterior density $\Post$,
where $\Prior$ is the prior on $\theta$, $\Like$ is the data
likelihood conditional on $\theta$, and $\Ly$ is the marginal
likelihood of the data.  Therefore,
\begin{align}
  \label{eq:Post}
  \Post=\frac{\LikePrior}{\Ly}=\frac{\Dy}{\Ly}
\end{align}
where $\Dy$ is the joint
density of the data and the parameters (of the unnormalized posterior
density).  In a marketing research context, under the conditional
independence assumption, the likelihood can be factored as
\begin{align}
\label{eq:dataLike}
f(y|\theta)=\prod_{i=1}^Nf_i\left(y_i|\beta_i,\alpha\right)
\end{align} 
where $i$ indexes households.\footnote{For brevity, we use the term
  ``household'' to describe any heterogeneous unit.}  Each $y_i$ is a vector of observed data, each $\beta_i$ is a vector of
heterogeneous parameters, and $\alpha$ is a vector of homogeneous population-level
parameters.  The $\beta_i$ are distributed across the population of
households according to a mixing distribution $\pi(\beta_i|\alpha)$,
which also serves as the prior on each $\beta_i$.  The elements of $\alpha$
may influence either the household-level data likelihoods, or the
mixing distribution (or both).  In this example, $\theta$ includes all
$\beta_1\mathellipsis\beta_N$ and all elements of $\alpha$.  The
prior itself can be factored as
\begin{align}
\label{eq:prior}
\pi(\theta)=\prod_{i=1}^N\pi_i(\beta_i|\alpha)\times\pi(\alpha).
\end{align} 
Let $\theta^*$ be the mode of
$\Dy$, which is also the mode of $\Post$, since $\Ly$ is a constant
that does not depend on $\theta$.  One will probably use
some kind of iterative numerical optimizer to find $\theta^*$, such
as a quasi-Newton line search or trust region algorithm. Define
$c_1=\mathcal{D}(\theta^*,y)$, and choose a proposal
distribution $\Gtheta$ that also has its mode at $\theta^*$. Define
$c_2=g(\theta^*)$, and define the function
\begin{align}
  \label{eq:PhiDef}
\Phitheta=\displaystyle\frac{\LikePrior \cdot c_2}{\Gtheta \cdot c_1}  
\end{align}
Through substitution and rearranging terms, we can write the target posterior density as
\begin{align}
  \label{eq:PostDefPhi}
  \Post=\Phitheta\cdot\Gtheta\cdot\frac{c_1}{c_2\cdot\Ly}
\end{align}

An important restriction on the choice of $g(\theta)$ is that the inequality $0\leq\Phi(\theta|y)\leq 1$ must hold, at least for any
$\theta$ with a non-negligible posterior density. We discuss this restriction, along with the choice of $\Gtheta$, in more detail a little later.

Next, let $u|\theta,y$ be an auxiliary variable that is distributed uniformly on
$\left( 0,\displaystyle\frac{\Phitheta}{\pi(\theta|y)}\right)$, so that
$p(u|\theta, y)=\displaystyle\frac{\pi(\theta|y)}{\Phitheta}=\frac{c_1}{c_2\Ly}g(\theta)$.  Then construct a joint density of
$\theta|y$ and $u|\theta, y$, where\begin{align}\label{eq:joint}
 p(\theta,u|y)&=\frac{\pi(\theta|y)}{\Phitheta}\mathbb{1}\left[ u<\Phitheta\right] 
\end{align}
By integrating Equation \ref{eq:joint} over $u$, the marginal density of $\theta|y$ is
\begin{align}
  \label{eq:margThetaY}
  p(\theta|y)&=\frac{\pi(\theta|y)}{\Phitheta}\int_0^{\Phitheta}~du =\pi(\theta|y)
\end{align}
Simulating
from $p(\theta|y)$ is now equivalent to simulating from the target
posterior $\pi(\theta|y)$.

Using Equations \ref{eq:PostDefPhi} and \ref{eq:joint}, the marginal density of $u|y$ is
\begin{align}
  \label{eq:margU}
p(u|y)&=\int_{\theta} \frac{\pi(\theta|y)}{\Phi(\theta|y)}\mathbb{1}\left[
  u<\Phi(\theta|y)\right]  d\theta\\
&=\frac{c_1}{c_2\Ly}\int_{\theta} \mathbb{1}\left[
  u<\Phi(\theta|y)\right] g(\theta)~d\theta\\
&=\frac{c_1}{c_2\Ly}q(u)\label{eq:marg_uy}
\end{align}
where $q(u) =\int_{\theta} \mathbb{1}\left[
  u<\Phi(\theta|y)\right] g(\theta)~d\theta$ .  This $q(u)$ function
is the probability that any candidate draw from $\Gtheta$ will satisfy
$\Phitheta > u$.  The sampler comes from recognizing that $p(\theta,u|y)$ can
be written differently from, but equivalently to, Equation \ref{eq:joint}.
\begin{align}
  \label{eq:joint2}
  p(\theta,u|y)&=p(\theta|u,y)~p(u|y)
\end{align}
The method involves sampling a $u$ from an approximation to $p(u|y)$, and then sampling from
$p(\theta|u,y)$.  Using the definitions in Equations \ref{eq:PhiDef},
\ref{eq:PostDefPhi}, and \ref{eq:joint}, we get
\begin{align}
  \label{eq:GDS_AR}
  p(\theta|u,y)&=\frac{p(\theta,u|y)}{p(u|y)}\\
&=\frac{c_1}{c_2\Ly p(u|y)}\mathbb{1}[u<\Phitheta]~g(\theta)
\end{align}
To sample \emph{directly} from $p(\theta,u|y)$, one
needs only to sample from $p(u|y)$ and then sample repeatedly from
$g(\theta)$ until $\Phi(\theta|y)>u$. The samples of $\theta$ form the
marginal distribution $p(\theta|y)$, and since sampling from $p(\theta|y)$
is equivalent to sampling from $\pi(\theta|y)$, they form an empirical
estimate of the target posterior density.

\subsection{Implementation}
But how does one simulate from $p(u|y)$?  In Equation \ref{eq:margU},
we see that $p(u|y)$ is proportional to the function $q(u)$.
\citet{WalkerLaud2011} sample from a similar kind of density by first taking $M$ proposal draws from the prior to construct
an empirical approximation to $q(u)$, and then approximating that 
continuous density using Bernstein polynomials.  However, in
high-dimensional models, this approximation tends to be a poor one at
the endpoints, even with an extremely large number of Bernstein polynomial
components.

Our approach is similar in that we effectively trace out an empirical
approximation to $q(u)$ by repeatedly sampling from $g(\theta)$, and
computing $\Phitheta$ for each of those proposal draws.  To avoid the
endpoint problem in the \citeauthor{WalkerLaud2011} method, we instead sample a 
transformed variable
$v=-\log u$.  Applying a change of variables,
$q_v(v)=q(u)\exp(-v)$.  With $q_v(v)$ denoting the ``true'' CDF of $v$,
let $\widehat{q}_v(v)$ be the empirical CDF of $v$ after taking $M$ proposal draws from
$\Gtheta$, and ordering the proposals
$0<v_1<v_2<\mathellipsis<v_M<\infty$. To be clear,
$\widehat{q}_v(v)$ is the proportion of samples that are \emph{strictly}
less than $v$.  As $M$ becomes large, the empirical approximation
becomes more accurate.

Because $\widehat{q}_v(v)$ is discrete, we can sample from a
density proportional to $q(u)\exp(-v)$ by partitioning the domain into
$M+1$ segments with the break point of each partition at each $v_i$.  The probability of
sampling a new $v$ that falls between $v_i$ and $v_{i+1}$ is now
\begin{align}
\varpi_i=\widehat{q}_v(v)\left[\exp(-v_i)-\exp(-v_{i+1})\right],
\end{align}
so we
can sample an interval bounded by $v_i$ and $v_{i+1}$ from a multinomial density
with weights proportional to $\varpi_i$.  Once we have the $i$ that
corresponds to that interval, we
can sample the continuous $v$ by sampling $\epsilon$  from a standard
exponential density, truncated on the right at $v_{i+1}-v_i$, and
setting $v=v_i + \epsilon$.  Thus, we can sample $v$ by first 
sampling $i$ with weight $\varpi_i$, then sampling a standard uniform
random variable $\eta$, and finally setting
\begin{align}
v=v_i-\log\left[1-\eta\left(1-\exp(v_i-v_{i+1}\right)\right].
\end{align}

To sample $R$ independent draws from the target posterior, we need $R$ ``threshold''
draws of $v$.  Then, for each $v$, we repeatedly sample from $\Gtheta$ until  $-\log(\Phitheta)< v$.  Once we have a $\theta$
that meets this criterion, we save it as a valid sample from
$\pi(\theta|y)$.  The complete algorithm is summarized as Algorithm \ref{alg:GDS}.

\begin{algorithm}
\caption{Algorithm to collect $R$ samples from $\Post$}
\label{alg:GDS}
\begin{algorithmic}[1]
\STATE{$R\leftarrow$ number of required samples from $\Post$}
\STATE{$M\leftarrow$ number
  of proposal draws for estimating $\widehat{q}_v(v)$.}
\STATE{$\theta^*\leftarrow$ mode of $\Dy$}
\STATE{$c_1\leftarrow\mathcal{D}(\theta^*,y)$}
\STATE{{\tt FLAG}$\leftarrow$ {\tt TRUE}}
\WHILE{{\tt FLAG}}
\STATE{Choose new proposal distribution $\Gtheta$}
\STATE{{\tt FLAG}$\leftarrow${\tt FALSE}}
\STATE{$c_2\leftarrow g(\theta^*)$.}
\FOR{$m:=1$ \TO $M$}
\STATE{Sample $\theta_m \sim \Gtheta$.}
\STATE{$\log\Phi(\theta_m|y)\leftarrow
  \log\mathcal{D}(\theta_m,y)-\log g(\theta_m)-\log c_1 + \log c_2$.}
\STATE{$v_m=-\log\Phi(\theta_m|y)$}
\IF{$\log\Phi(\theta_m|y)>0$}
\STATE {{\tt FLAG}$\leftarrow$ {\tt TRUE}}
\STATE{{\bf break}}
\ENDIF
\ENDFOR
\ENDWHILE
\STATE{Reorder elements of $v$, so
  $0<v_1<v_2<\mathellipsis <v_M<\infty$. Define $v_{M+1}:=\infty$}
\FOR{$i:=1$ \TO $M$}
 \STATE{$\widehat{q}_v(v_i)\leftarrow\sum_{j=1}^M\mathbb{1}\left[v_j<v_i\right]$.}
\STATE{ $\varpi_i\leftarrow \widehat{q}_v(v_i)\left[\exp(-v_i)-\exp(-v_{i+1})\right]$.}
\ENDFOR
\FOR{$r=1$ \TO $R$}
\STATE{Sample $j\sim$ Multinomial$(\varpi_1\mathellipsis\varpi_M)$.}
\STATE{Sample $\eta\sim$ Uniform(0,1).}
\STATE{$v^*\leftarrow v_j-\log\left[1-\eta\left( 1-\exp\left( v_j-v_{j+1}\right)\right)\right]$.}
\STATE{$p\leftarrow 0$}
\STATE{$n_r\leftarrow 0$.  }\COMMENT{Counter for number of proposals}
\WHILE{$p > v^*$}
\STATE{Sample $\theta_r\sim\Gtheta$.}
\STATE{$p\leftarrow -\log\Phi\left(\theta_r|y\right)$.}
\STATE{$n_r\leftarrow n_r+1$.}
\ENDWHILE
\ENDFOR
\RETURN{$\theta_1\mathellipsis \theta_R$ (plus $n_1\mathellipsis n_R$ and $v_1\mathellipsis v_M$
  if computing a marginal likelihood).}
\end{algorithmic}
\end{algorithm}

\subsection{The proposal distribution}
The only restriction on $\Gtheta$ is
that the inequality $0\leq\Phi(\theta|y)\leq 1$ must hold, at least for any
$\theta$ with a non-negligible posterior density.  Because $v>0$, we
must have $u<1$.  Thus, any $\theta$ for which $\Phitheta>1$ would
always be accepted, no matter how small $\pi(\theta|Y)$ might be. By
construction, $\Phi(\theta^*|y)=1$, meaning that no candidate $\theta$
will have a higher acceptance probability than the $\theta$ with the
highest posterior density.  This is an intuitively appealing
property. 

 In principle, it is up to the researcher to choose
$\Gtheta$, and some choices may be more efficient than others.  We have found that a multivariate
normal (MVN) proposal distribution, with mean at $\theta^*$, works well for
the kinds of 
continuous posterior densities that marketing researchers typically
encounter.  The MVN density, with a covariance matrix
equal to the negative inverse Hessian of the log posterior at $\theta^*$, is an
asymptotic approximation (specifically, a second-order Taylor series) to the posterior density itself
\citep[sec. 5.2]{CarlinLouis2000}.  By multiplying that covariance
matrix by a scaling constant $s$, we can derive a
proposal distribution that has the general shape of the target posterior
near its mode.  That proposal distribution will be valid as long as $s$ is large enough so that
$\Phitheta$ is between 0 and 1 for any plausible value of $\theta$,
and that the mode of $\Gtheta$ is at $\theta^*$.

\begin{figure}[tb]
  \centering
  \includegraphics[trim=0 15 0 12,clip]{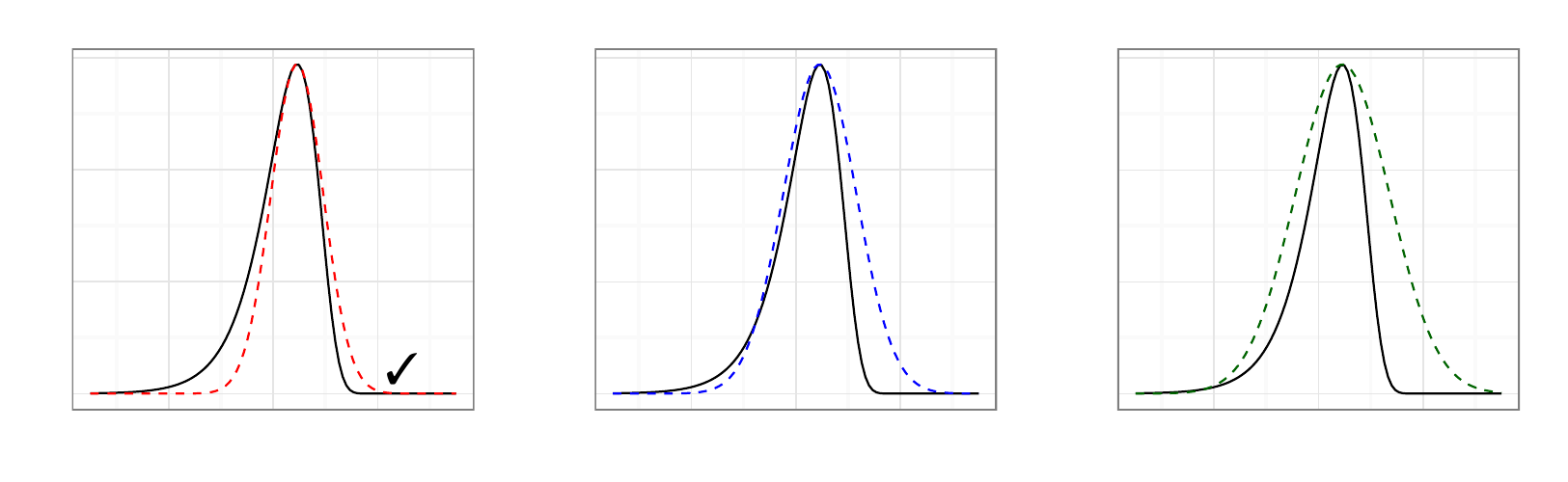}
  \caption{A "target" posterior density (solid line, same in all panels), and three scaled normal densities (dotted lines, increasing in covariance from left to right).}
  \label{fig:scaledNorm}
\end{figure}

We illustrate the idea of scaling the proposal density in Figure
\ref{fig:scaledNorm}.  The solid line (the same in all three panels)
is a ``target'' posterior density.  The dotted lines plot potential normal
proposal densities, multiplied by the corresponding $c_2/c_1$ ratio.
The covariance of the proposal density in the left panel is the
negative inverse Hessian of the log posterior density.  Samples from the left tail of the posterior distribution will
have $\Phitheta>1$.  In the middle and right panels, the covariance is
the same as in the left panel, but multiplied by 1.4 and 1.8,
respectively.  As the covariance increases, more of the posterior
support will have $\Phitheta\leq 1$.  

It is possible that $\Gtheta$ could under-sample values of
$\theta$ in the tails of the posterior.  However, if $M$ is
sufficiently large, and $\Phitheta\leq 1$ for all $M$ proposals, then
it is unlikely that we would see $\Phitheta > 1$ in the rejection
sampling phase of the algorithm.  If that were to happen, we can stop,
rescale, and try again.  Any values of $\theta$ that we might miss
would have such low posterior density that there would be little
meaningful effect on inferences we might make from the output.

We believe that the potential cost from under-sampling
the tails is dwarfed by our method's relative computational advantage, we
recognize that there may be some applications for which sampling
extreme values of $\theta$ may be important.  In that case, this
may not be the best estimation method for those applications.  Otherwise,
there is nothing special about using the MVN for $\Gtheta$.  It is
straightforward to implement with manual adaptive selection of $s$.
This is similar, in spirit, to the concept of tuning a Metropolis-Hastings algorithm.  Heavier-tailed proposals,
such as the multivariate-t distribution, can fail because of high kurtosis
at the mode. 

\subsection{Comparison to Other Methods}\label{sec:compare}
\subsubsection{Rejection Sampling}
At first glance, our method looks quite similar to standard rejection
sampling.  With rejection sampling, one repeatedly samples both a threshold value from
a standard uniform distribution ($p(u)=1$), and a proposal from $g(\theta)$ until $\Post/\Gtheta\geq Ku$, where $K$ is a
positive constant.  This is different from our method, for which the
threshold values are sampled from a posterior $p(u|y)$, and $K$ is
specifically defined as the ratio of modal densities of the posterior
to the proposal.  The advantages that rejection sampling has
over our approach is that the distribution of $u$ is exact, and that the
proposal density does not have to dominate the target density for all
values of $\theta$.  However, for
any model with more than a few dimensions, the critical ratio can be
extremely small for even small deviations of $\theta$ away from the
mode.  Thus, the acceptance probabilities
are virtually nil.  In contrast, we accept a discrete
approximation of $p(u|y)$ in exchange for higher
acceptance probabilities.

\subsubsection{Direct Sampling}
\citet{WalkerLaud2011} introduced and demonstrated the merits
of a non-MCMC approach called Direct Sampling for conducting
Bayesian inference.  Like our method, Direct Sampling removes the need to concern
oneself with issues like chain convergence and autocorrelation, and generates independent samples
from a target posterior distribution \emph{in parallel}.
\citet{WalkerLaud2011} also prove that the sample acceptance probabilities using Direct
Sampling are better than those from standard 
rejection algorithms.  Put simply, for many common Bayesian models, they demonstrate 
improvement over MCMC in terms of efficiency, resource demands and
ease of implementation.  

However, Direct Sampling suffers from some important shortcomings that limit its broad
applicability. One is the failure to
separate the specification of the prior from the specifics of the
estimation algorithm.  Another is an inability to generate accepted draws for
even moderately sized problems; the largest number of parameters that
Walker et al. consider is 10.  Our method allows us to conduct full Bayesian inference on hierarchical models in high dimensions,
with or without conjugacy, without MCMC.  

Although we share some important features with Direct Sampling, our
method differs in several respects.  While Direct Sampling focuses on
the shape of the data likelihood alone, we are concerned with the
characteristics of the entire posterior density. Our method bypasses
the need for Bernstein polynomial approximations, which are integral
to the Direct Sampling algorithm. Finally, while Direct Sampling takes
proposal draws from the prior (which may conflict with the data), our
method samples proposals from a separate density that is ideally a
good approximation to the target posterior density itself.

\subsubsection{Markov chain Monte Carlo}

We have already mentioned the key advantage of our method over traditional MCMC:
generating independent samples that can be collected in parallel.
We do not need to be concerned with issues like
autocorrelation, convergence of estimation chains, and so
forth. Without delving into a discussion of all possible variations
and improvements to MCMC that have been proposed in the last few
decades, there have been some attempts to parallelize MCMC that
deserve some mention.  For a deeper analysis, see \citet{SuchardWang2010}.

It is possible to run multiple independent MCMC chains that start from
different starting points.  Once all of those chains have converged to
the posterior distribution, we can estimate the posterior by combining samples from all of the
chains.  The numerical efficiency of that approximation should be
higher than if we used only one chain, because there should be no
correlation between samples collected in different chains.  However,
each chain still needs to converge to the posterior independently, and
only after that convergence could we start collecting samples.  If it
takes a long time for one chain to converge, it will take at least
that long for all chains to converge.  Thus, the potential for
parallelization is much greater for our method than for MCMC.

Another approach to parallelization is to exploit parallel processing
power for individual steps in an algorithm.  One example is a parallel implementation of a multivariate slice
sampler (MSS), as in \citet{TibbitsHaran2011}.   The benefits of parallelizing the MSS
come from parallel evaluation of the target density at each of the vertices of the
slice, and from more efficient use of resources to execute
linear algebra operations (e.g, Cholesky decompositions). But the MSS
itself remains a Markovian algorithm, and thus will still generate dependent
draws.  Using parallel technology to generate a single draw from a distribution is not the
same as generating all of the required draws themselves in parallel.
The sampling steps of our method can be run in their
\emph{entirety} in parallel.  

Another attractive feature of our method is that the model is fully specified by the log
posterior, and possibly its gradient and Hessian.  The tools that we
discuss in Section \ref{sec:scalability} are components of a reusable
infrastructure.  Only the function that returns the log posterior
changes from model to model.  This feature is unlike a blockwise Gibbs sampler,
for which we need to derive and implement a set of conditional
densities for each model.  A small change in the model specification
can require a substantial change in the sampling strategy.  For
example, a change to a prior might mean that the sampler is no longer
conditionally conjugate.  So while it might be possible to construct
a highly efficient Gibbs sampler for a particular model (e.g.,
through blocking, data augmentation, or parameter transformation),
there can be considerable upfront investment in doing so.

\section{Examples}\label{sec:empirical}

We now provide three examples of our method in action.  In the first
example, we simulate data from a basic, conditionally conjugate model,
and compare the marginal posterior distributions that were generated
by our method with those from a Gibbs sampler.  The second example is
a non-conjugate hierarchical model of store-level retail sales.  In
that example, we compare estimates from our method with those from
Hamiltonian Monte Carlo (HMC), using the Stan software package
\citep{STAN}. For these first two examples, the MCMC methods are
efficient, so it is likely that they generate good estimates of the
true posterior densities.  Thus, we can use those estimates as
benchmarks against which we can assess the accuracy of our method.
The third example is a more complicated model for which MCMC performs
poorly. We use this example to not only assess the accuracy of our
method (for those parameters for which we think the MCMC estimates are
reasonable), but to illustrate some of the computational problems that
are inherent in MCMC methods.  In all of our examples, we implemented
our method using the \pkg{bayesGDS} package \citep{R_bayesGDS} for \proglang{R}.

\subsection{Simulated data, conditionally conjugate model}
In our first example, we simulated $T=10$ observations for each of $N=1500$ heterogeneous units.
Each observation for unit $i=1\mathellipsis N$ is a sample from a
normal distribution, with mean $\theta_i$ and standard deviation
$\sigma=2$.  The $\theta_i$ are normally-distributed across the
population, with mean $\mu=-1$ and standard deviation $\tau=3$. We
place uniform priors on $\mu$, $\log \sigma$ and $\tau$.  There are
1,503 parameters in this model.

This model allows for a conditionally conjugate Gibbs sampler; the
steps are described in \citet[][Section 11.7]{GCSR2004}.  In this
case, the Gibbs sampler is sufficiently fast and efficient, so we have confidence
that it does indeed sample from the correct posterior
distributions.  The Gibbs sampler was run
for 2,000 iterations, including a 1,000-iteration burn-in.  The process took about five
minutes to complete.  We then collected 360 independent samples using our method, after estimating
$q_v(v)$ with $M=70,000$ proposals, and applying a scaling factor on
the inverse Hessian of 1.02.  To get those 360 proposals, we needed
381,507 proposals.  Using a single core of a 2014-vintage Mac Pro,
this process would take about 23 minutes.  However, each draw can be
collected in parallel. It took about five minutes to collect all 360
samples when using 12 processing cores.  The mean
number of proposals for each posterior sample was 1,060 (an acceptance
rate of .0009), but the median was only 29.  Ten of the 360 samples
required more than 10,000 proposals. 

\begin{figure}[tb]
  \centering
  \includegraphics[trim=0 12 0 12,clip]{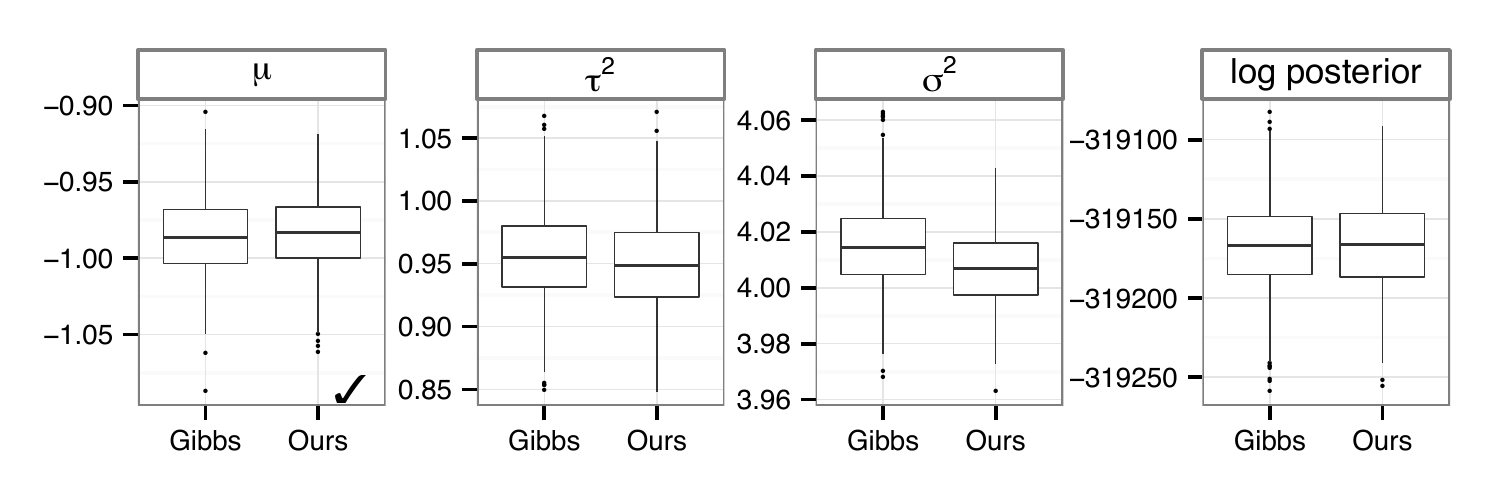}
  \caption{Quantiles of posterior densities from hierarchical model with conditional conjugacy, as estimated using our method, and with a Gibbs sampler.}
  \label{fig:conjPlot}
\end{figure}

Figure \ref{fig:conjPlot} shows the quantiles of samples in the
estimated marginal posterior densities of the population-level
parameters $\mu$, $\tau^2$, and $\sigma^2$, as well as the log of the
unnormalized posterior density. 
We can see that both methods generate
effectively the same estimated posterior distributions.

\subsection{Hierarchical model without conditional conjugacy}
In the second example, we model weekly sales of sliced cheese in 88 retail
stores.  The data are available in the \pkg{bayesm} package for
\proglang{R} \citep{R_bayesm}, and
were used in \citet{BoatwrightMcCulloch1999}.  Under this model,
mean sales for store $i$ in week $t$ has a gamma distribution with mean $\lambda_{it}$
and shape $r_i$.  The mean is a log-linear function of price, and
the percentage of ``all category volume'' on display in that week.
\begin{align}
  \label{eq:2}
  \log \lambda_{it}=\beta_{i1}+\beta_{i2}\log\text{PRICE}_{it}+\beta_{i3}\text{DISP}_{it}
\end{align}
The prior on each $r_i$ is a half-Cauchy distribution with a
scale parameter of 5, and the prior on each $\beta_i$ is MVN with mean
$\mu$ and covariance $\Omega$.  The hyperparameters $\mu$ and $\Omega$ have weakly informative MVN and inverse Wishart
priors, respectively. There are 361 parameters in this model.

This model does not allow for a conditionally conjugate Gibbs
sampler.  Instead, we
use Hamiltonian Monte Carlo (HMC, \citealt{DuaneKennedy1987,Neal2011}), which
uses the gradient of the log posterior to simulate a dynamic physical
system.  We selected HMC mainly because it is known
to generate successive samples that are less serially correlated
than draws that one might sample using other MCMC methods. 

\begin{figure}[tb]
  \centering
  \includegraphics[trim=0 10 0 10,clip]{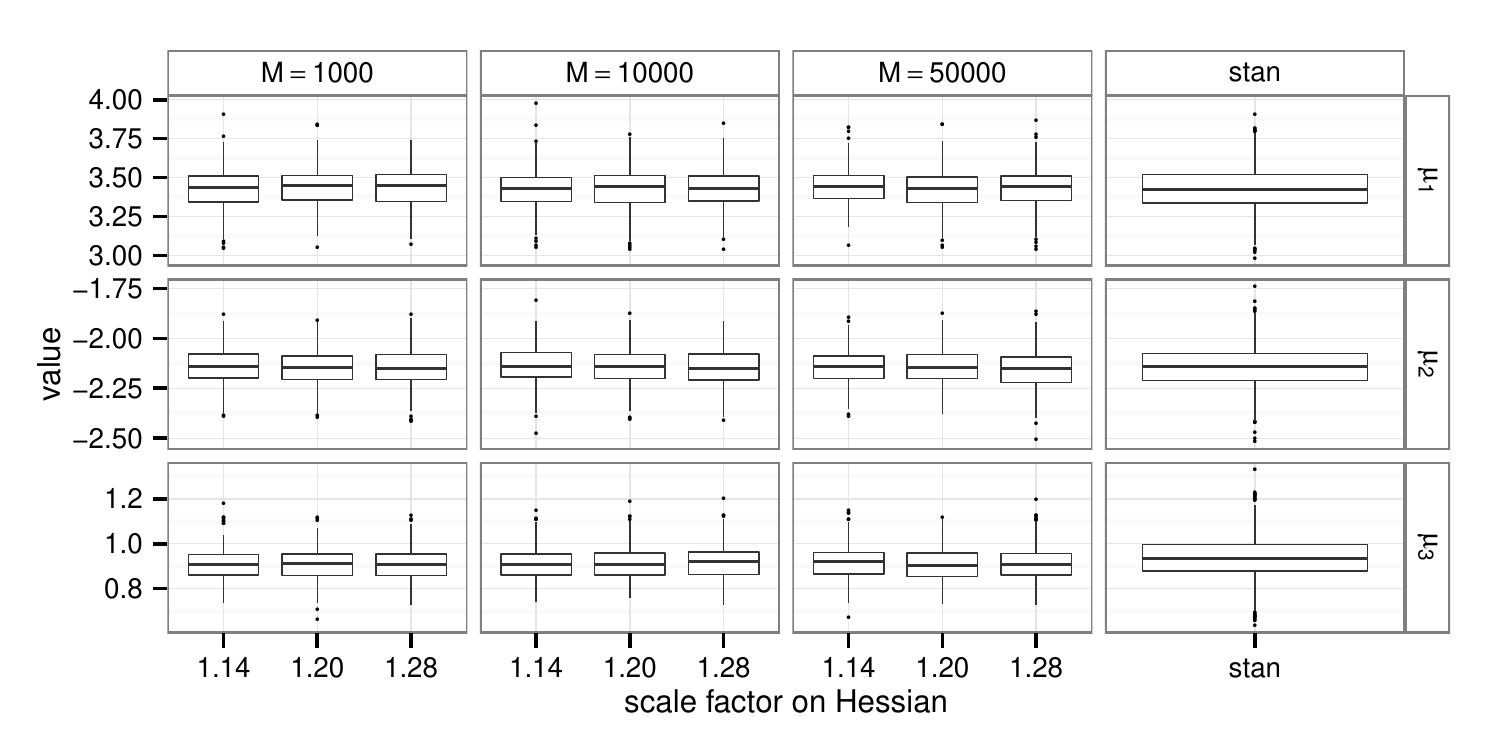}
  \caption{Estimated posterior densities of $\mu$ for the sliced cheese example.}
  \label{fig:cheese}
\end{figure}

We implemented HMC using the \pkg{Stan} software package \citep{STAN}.  We
ran 5 parallel chains for 800 iterations each, and discarding the
first half of the draws.  The chains appeared to display little
autocorrelation, so we have confidence that the HMC samples form a
good estimate of the true posterior distributions.  We then estimated the model using our method with different numbers of proposal draws to estimate $q_v(v)$, and different
scale factors on the covariance of the proposal density. In Figure
\ref{fig:cheese}, we compare the estimates densities for elements of
$\mu$, the baseline and marginal effects on sales.  We see that even
with a relatively coarse estimate of $q_v(v)$, and an overly diffuse
proposal density, our method generates estimates of the posterior
densities that are not only comparable with each other, but also
comparable those generated by Stan.  The acceptance rates across
different runs using our method was .0003.  It took about 1.4 seconds
to sample and evaluate a block of 1,000 proposals on a single
processing core.

\subsection{Model with weakly identified parameters}
Our third example concerns a non-conjugate
heterogeneous hierarchical model, in which
some parameters are only weakly identified.  The
data structure is described in \citet{ManchandaRossiChintagunta2004};
we use the simulated data that are available in the \pkg{bayesm}
package for \proglang{R} \citep{R_bayesm}.  In this dataset, for each of 1,000 physicians, we observe weekly prescription
counts for a single drug ($y_{it}$), weekly counts of sales visits from
representatives of the drug manufacturer ($x_{it}$), and some
time-invariant demographic information ($z_i$).  Although one could model the
purchase data as conditional on the sales visits,
\citeauthor{ManchandaRossiChintagunta2004} argue that the rate of
these contacts is determined endogenously, so that physicians who are
expected to write more prescriptions, or who are more responsive to
the sales visits, will receive visits more often. 

In this model, $y_{it}$ is
a random variable that follows a negative binomial distribution with
shape $r$ and mean $\mu_{it}$, and $x_{it}$ is a
Poisson-distributed random variable with mean $\eta_i$.  The expected
number of prescriptions per week depends in the number of sales visits, so we let $\log\mu_{it}=\beta_{i0}+\beta_{i1}x_{it}$, where
$\beta_i$ is a vector of heterogeneous coefficients.  We then
model the contact rate so it depends on the physician-specific
propensities to purchase, so
$\log\eta_i=\gamma_0+\gamma_1\beta_{i0}+\gamma_2\beta_{i1}$.  Define
$z_i$ as a vector of four physician-specific demographics (including an
intercept), and define $\Delta$ as
a $2\times 4$ matrix of population-level coefficients.  The mixing distribution
for $\beta$ (i.e., the prior on each $\beta_i$), is MVN with mean
$\Delta'z_i$ and covariance $V$.  We place weakly informative MVN
priors on $\gamma$ and the rows of $\Delta$, an inverse Wishart
prior on $V$, and a gamma prior on $r$.  There are
2,015 distinct parameters to be estimated.
This model differs slightly from the one in
\citet{ManchandaRossiChintagunta2004} only in that $\eta_i$ depends
only on expected sales in the current period, and not the long-term
trend.  We made this change to make it easier to run the baseline MCMC
algorithm.

As before, our baseline estimation algorithm is HMC, but instead of
using Stan, we use the ``double averaging'' method
to adaptively scale the step size \citep{HoffmanGelman2014}, and we set
the expected path length to 16.\footnote{Shorter path lengths were less efficient,
and longer ones frequently jumped so far from the regions of high
posterior mass that the computation of the log posterior would
underflow.  We had the same
problem with the No U-Turn Sampler \citep{HoffmanGelman2014}, whether
using Stan, or coding the algorithm ourselves.  The HMC
extensions in \citet{GirolamiCalderhead2011} are inappropriate for
this problem because the Hessian is not guaranteed to be positive
definite for all parameter values.}  By implementing HMC ourselves, we
can use the same computer code to compute the log posterior, and its gradient, that we
use with our method.  This allows the two methods to compete on a
level playing field.

\begin{figure}[tb]
   \centering
   \includegraphics[scale=1,trim=0 10 0 12,clip]{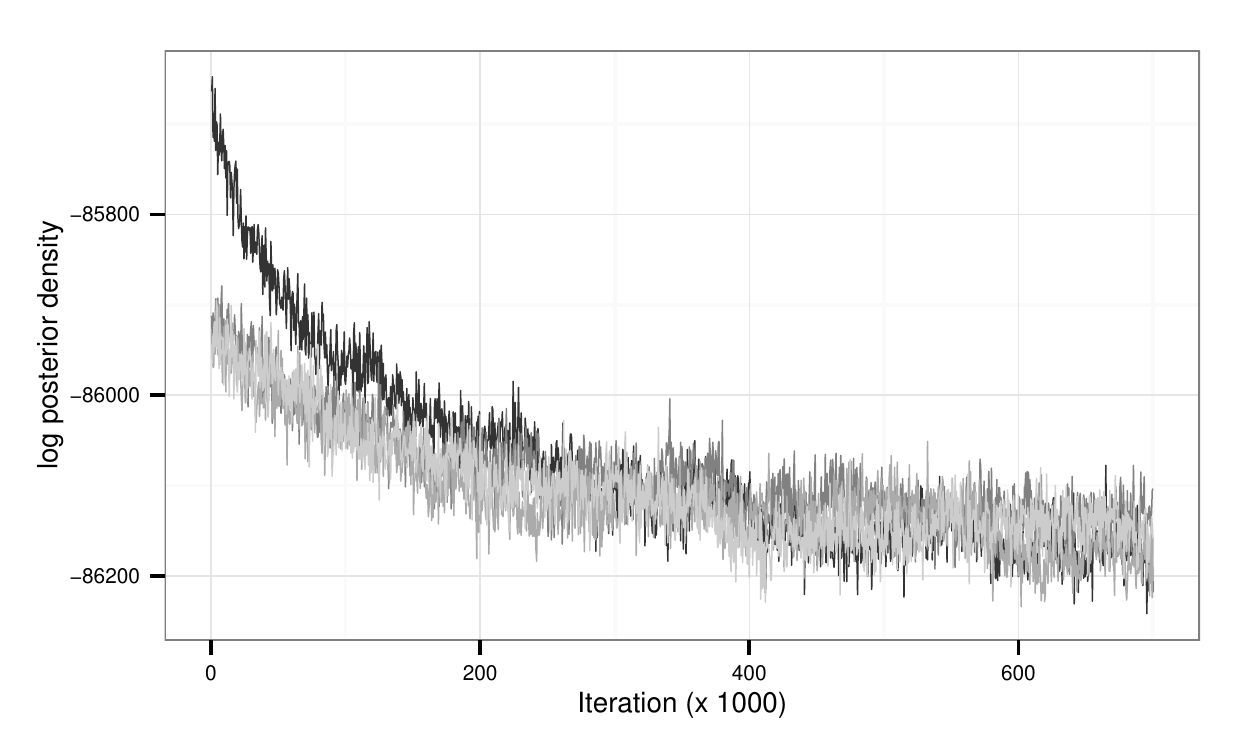}
  \caption{HMC trace plot of log posterior density for four chains. One
     chain was started at the mode, and the others started at random
     values. Every 500th iteration is plotted.}
   \label{fig:logpost_trace}
 \end{figure}

\begin{figure}[tb]
   \centering
   \includegraphics[scale=1,trim=0 5 0 10,clip]{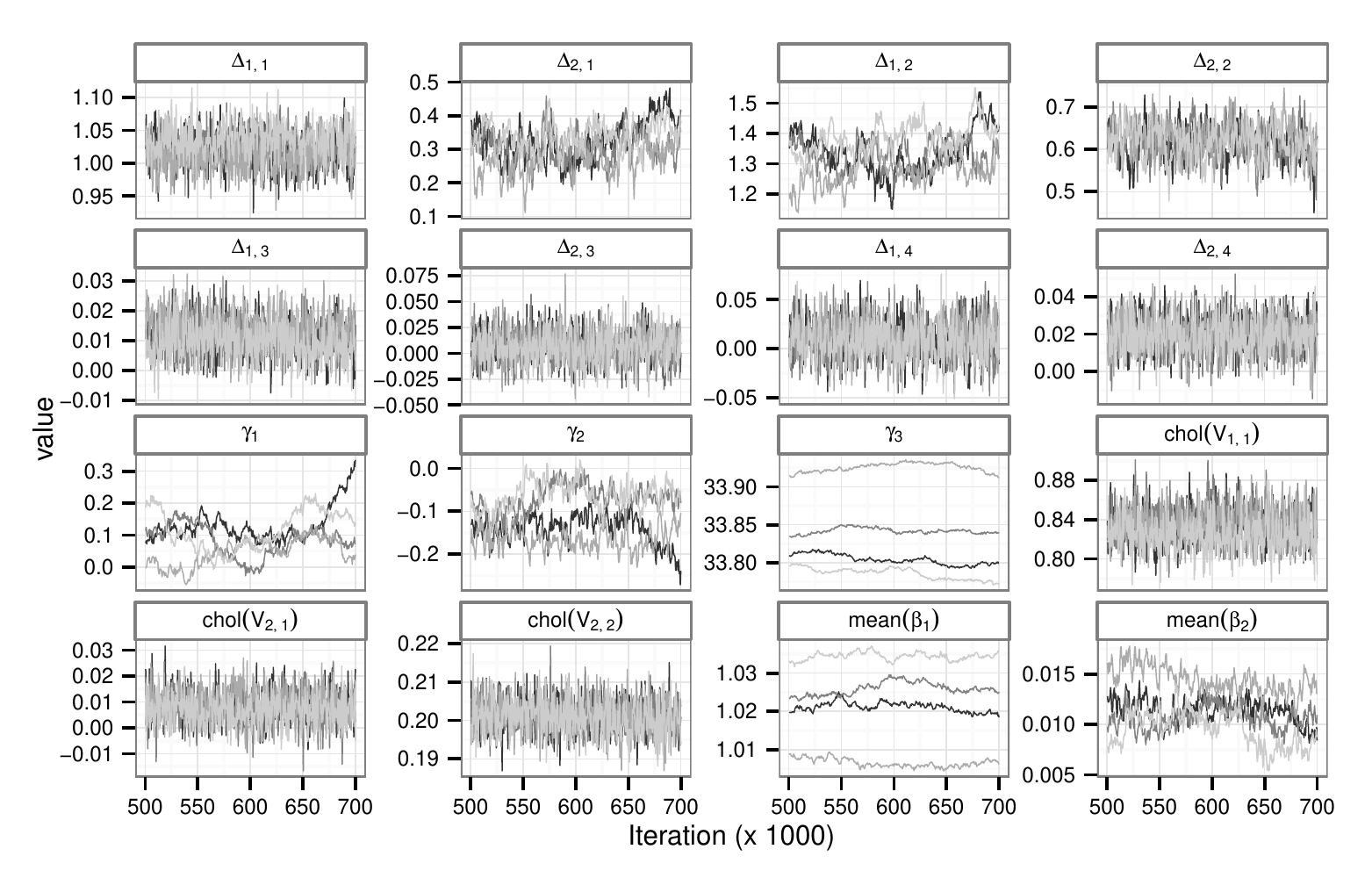}
   \caption{HMC trace plot for value of population-level parameters in hierarchical model example, starting at Iteration 500,000.  Every 500th iteration is plotted.}
   \label{fig:marg_trace}
\end{figure}

We ran four
independent HMC chains for 700,000 iterations each, during a period
of more than three weeks.  Searching for the posterior mode is considered, in
general, to be ``good practice'' for Bayesian inference, and
especially with MCMC; see Step 1 of the ``Recommended Strategy for Posterior
Simulation'' in Section 11.10 of \citet{GCSR2004}.  Finding the mode
of the log posterior is the first step of our method anyway, so we initialized
one chain there, and the other three at randomly-selected starting
values.  Figure \ref{fig:logpost_trace} is a
trace plots of the log posterior density. The chains begin to approach each other only after about 500,000
iterations.  The panels in Figure \ref{fig:marg_trace}, are trace plots of
the population-level parameters.  Some
parameter chains appear to have converged to each other, with little
autocorrelation, but others seem to make no
progress at all. Table \ref{tab:effSize}, summarizes the effective
sample sizes for estimates of the marginal posterior distributions of
population-level parameters, using the final 100,000 samples of the
HMC chains.  Many of these parameters may require more than a
million additional iterations to achieve an effective sample size
large enough to make reasonable inferences.

\begin{table}[ht]
\centering
\begin{tabular}{crrrrm{1cm}crrrr}
&  \multicolumn{4}{c}{Chain}&&&\multicolumn{4}{c}{Chain}\\
 & 1 & 2 & 3 & 4 &&&1&2&3&4\\ 
  \hline
  $\Delta_{1,1}$ & 295 & 250 & 266 & 267 &&  $\gamma_1$ & 3 & 4 & 6 & 2 \\ 
 $\Delta_{1,2}$ & 16 & 24 & 27 & 29 && $\gamma_2 $& 10 & 33 & 22 & 22 \\ 
 $\Delta_{1,3}$ & 6 & 12 & 26 & 14 && $\gamma_3$ & 3 & 15 & 3 & 3 \\ 
  $\Delta_{1,4}$ & 51 & 43 & 43 & 60 &&$\Chol(V)_{1,1}$ & 553 & 513 & 500 & 511 \\ 
  $\Delta_{2,1}$ & 4703 & 5933 & 5981 & 1444 &&  $\Chol(V)_{2,1}$ & 7844 & 18761 & 13796 & 9852 \\ 
  $\Delta_{2,2}$ & 507 & 573 & 526 & 538 &&  $\Chol(V)_{2,2}$ & 530 & 523 & 517 & 477 \\ 
  $\Delta_{2,3}$ & 407 & 445 & 400 & 407 &&  mean($\beta_{1i}$) & 6 & 5 & 12 & 12 \\ 
  $\Delta_{2,4}$ & 3708 & 4797 & 3680 & 4926 && mean($\beta_{2i}$) & 21 & 22 & 28 & 10 \\
   \hline
\end{tabular}
\caption{Effective sample sizes for estimates of marginal posterior
  distributions of population-level parameters, using the final
  100,000 samples of HMC chains.}
\label{tab:effSize}
\end{table}

The convergence problems are even worse when
we consider that each of the 16 steps in the path length for iteration requires one evaluation for both the
log posterior  and its gradient.  Using ``reverse mode'' 
automatic differentiation, which we discuss more in Section
\ref{sec:compute}, the time to evaluate the gradient is  about five times the time it takes to evaluate the log posterior,
regardless of the number of variables \citep{GriewankWalther2008}.
Therefore, each HMC iteration requires resources that equate to 96
evaluations of the posterior.  In other words, the computational cost
of 700,000 HMC iterations is equivalent to more than 67 million evaluations of
the log posterior.  And this assumes that 700,000 iterations were 
sufficient to collect enough samples from the true posterior.  

So how much more efficient is our sampling method?  We estimated the marginal density
$q_v(v)$ by taking $M=100,000$ proposal draws from an MVN distribution
with the mean at the posterior mode $\theta^*$, and the covariance matrix equal
the inverse of the Hessian at the mode, multiplied by a scaling
factor of $s=1.3$.  This value of $s$ is the smallest value for which
$0\leq\Phitheta\leq 1$ for all 100,000 samples from $\Gtheta$. We then collected 300
\emph{independent} samples \emph{in parallel} from the target posterior $\pi(\theta|y)$.
The median number of proposals required for each posterior sample was
just under 38,000, the total number of likelihood evaluations was
about 16.5 million, and average acceptance rate was $1.8\times
10^{-5}$.  

In absolute terms, the low acceptance rate appears to be unfavorable.
However, the total run time is much lower than for MCMC.  In our
implementation (using a single core on a 2014-vintage Apple Mac Pro),
the total time to compute the log posterior density of 1,000 proposal
samples is about 8.7 seconds.  The time to sample 1,000 proposals from
an MVN distribution, and compute the MVN densities for each, is about
.89 seconds.  Therefore, to collect and evaluate 16.5 million proposal
samples (to get 300 posterior draws) would take about 44 hours.  But this is for a single
processing core.  Using all 12 cores on our Mac Pro, the sampling time
is reduced to 220 minutes. We discuss the scalability of the component
steps of the algorithm in Section \ref{sec:scalability}.  

Access to more processing nodes could reduce this time even
further. It is the ability to collect posterior samples in parallel
that gives our method its greatest advantage over MCMC methods.  One
could run multiple MCMC chains in parallel, but this involves waiting until all of the
chains, individually, converge to the target posterior, before we
could begin collecting samples for inference.  Even then, there is no way to confirm that the chain has, in fact, converged.

\begin{figure}[tb]
  \centering
  \includegraphics[scale=1.1,trim=0 10 0 12,clip]{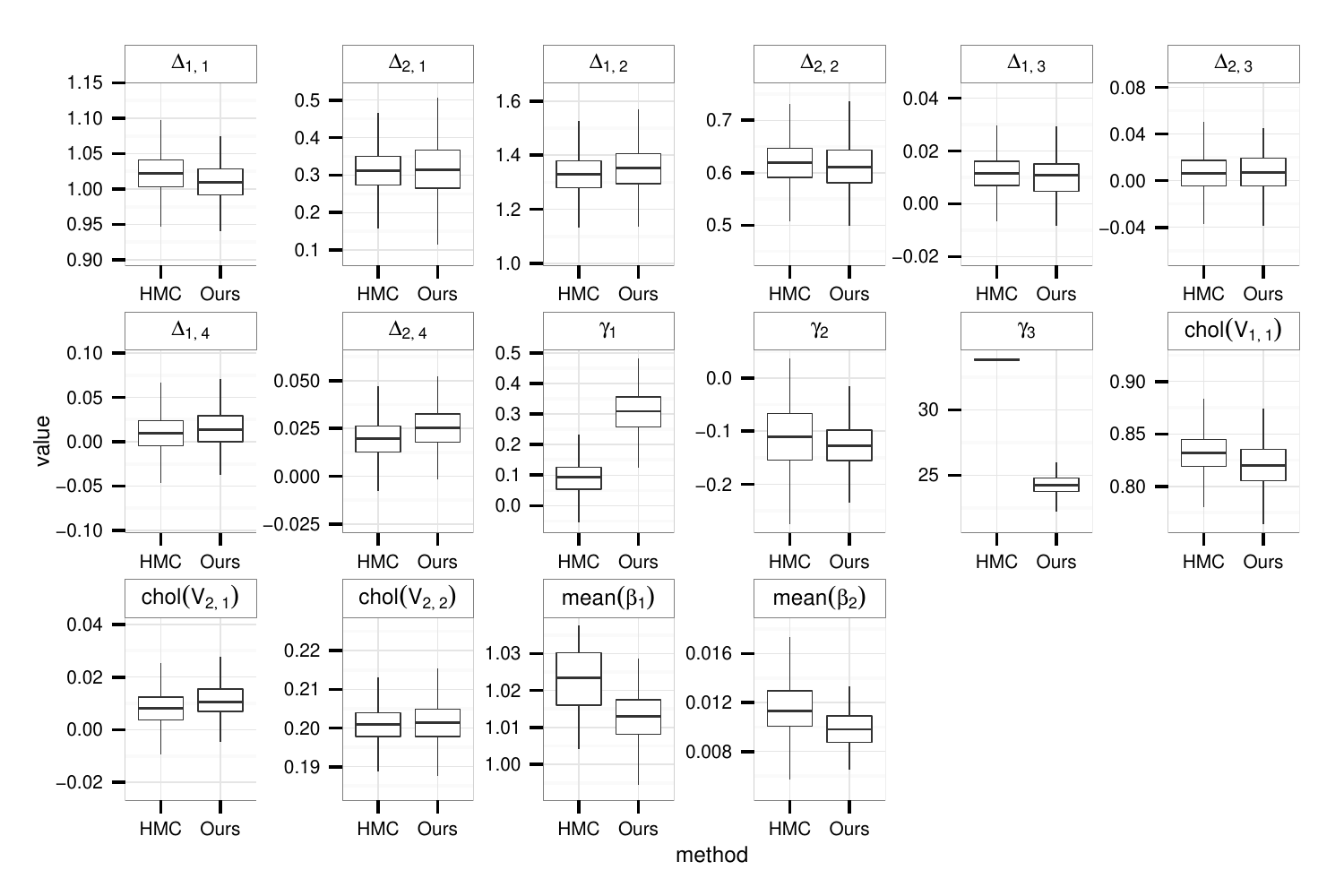}
  \caption{Estimated marginal distributions for population-level
    parameters using our method, and HMC.}
  \label{fig:comp_dens}
\end{figure}

We can draw inferences about the accuracy of our method by comparing the
estimated marginal densities to those that we get from HMC.  Note
that the HMC estimates are accurate only if all of the chains
 converge to the target density, and we have a large-enough
effective sample size.  This condition clearly does not hold, but it is sufficiently close for the majority of the
population-level parameters for us to use HMC samples as a baseline standard.
Figure \ref{fig:comp_dens} is a comparison of
the quantiles.  For the elements of $\Delta$ and the Cholesky decomposition of $V$, our
method's estimated distributions are close to the HMC estimates. For
other parameters, the convergence of the estimates is less clear.
However, the parameters for which the densities are not aligned are the same parameters for which there is high
autocorrelation, and little movement, in the HMC chains.  Thus, we
infer that our method compares with HMC well in terms of the marginal densities that
it generates, with substantially computational effort. 

\FloatBarrier
\section{Scalability and Sparsity}\label{sec:scalability}

The ability for our method to generate independent samples in
parallel already makes it an attractive alternative to MCMC.  In this
section, we present an argument in favor of our method's scalability.  Our
criterion for scalability is that the cost of running
the algorithm grows close to linearly in the number of
households.  Our analysis considers the fundamental computational
tasks involved:  computing the log posterior, its gradient
and its Hessian; computing the Cholesky decomposition of the Hessian;
and sampling from an MVN proposal distribution.   We will show
 that scalability can be achieved because, under the conditional
 independence assumption, the Hessian of the log posterior is sparse,
 with a predictable sparsity pattern.

\subsection{Computing Log Posteriors, Gradients and Hessians}\label{sec:compute}
Under the conditional independence assumption, the log posterior density is the sum of the logs of Equations
\ref{eq:dataLike} and \ref{eq:prior}, with a heterogeneous component
\begin{align}
  \label{eq:logPostHetero}
  \sum_i^N\log f_i\left(y_i|\beta_i,\alpha\right) + \log\pi_i(\beta_i|\alpha)
\end{align}
and a homogeneous component $\log \pi(\alpha)$. This homogeneous
component is the hyperprior on the population-level parameters, so its
computation does not depend on $N$, while each additional household
adds another element to the summation in Equation \ref{eq:logPostHetero}.
Therefore, computation of the log posterior grows linearly in $N$. In
the subsequent text, let $k$ be the number of elements in each
$\beta_i$ and let $p$ be the number of elements in $\alpha$. Using the
notation from Section \ref{sec:GDS}, $\theta$ is a vector that concatenates
all of the $\beta_i$ together, along with $\alpha$.

There are two reasons why we might need to compute the gradient and Hessian
of the log posterior, namely 
to use in a derivative-based optimization algorithm to find the
posterior mode, and for estimating the precision
matrix of an MVN proposal
distribution.\footnote{We do not require either derivative-based
  optimization or using MVN proposals, but these are most likely
  reasonable choices for differentiable, unimodal posteriors.}
Ideally, we would derive the gradient and Hessian analytically, and
write code to estimate it efficiently.  For complicated
models, the required effort for coding analytic gradients may not be
worthwhile. An alternative would be numerical approximation 
through finite differencing.  The fastest, yet least accurate, method
for finite differencing for gradients, using either forward or backward differences,
requires $Nk+p+1$ evaluations of the log posterior.  Since the
computational cost of the log posterior also grows linearly with $N$, computing
the gradient this way will grow quadratically in $N$. The cost of estimating a
Hessian using finite differencing grow even faster in $N$.  Also, if
the Hessian is estimated by taking finite differences of gradients,
and those gradients themselves were finite-differences, the
accumulated numerical error can be so large that the Hessian estimates
are useless.  

Instead, we can turn to automatic differentiation (AD, also sometimes known as algorithmic
differentiation).  A detailed explanation of AD is beyond the scope of
this paper, so we refer the reader to \citet{GriewankWalther2008}, or
Section 8.2 in \citet{NocedalWright2006}.  Put simply, AD treats a
function as a composite of subfunctions, and computes derivatives by
repeatedly applying the chain rule.  In practical terms, AD
involves coding the log posterior using a specialized numerical library that keeps
track of the derivatives of these subfunctions.  When we compile the
function that computes the log posterior, the AD library will ``automatically'' generate additional functions that
return the gradient, the Hessian, and even higher-order derivatives.\footnote{There are a number of established AD tools available for researchers to use for
many different programming environments.  For \proglang{C++}, we use \pkg{CppAD} \citep{CppAD2012}, although \pkg{ADOL-C}
\citep{WaltherGriewank2012} is also popular.  We call our \proglang{C++}
functions from \proglang{R} \citep{R_core} using
the \pkg{Rcpp} package \citep{R_Rcpp,R_RcppEigen}.  \pkg{CppAD} is
also available for \proglang{Python}.  \proglang{Matlab} users
have access to \pkg{ADMAT} \citep{ADMAT_2000}, among other options.}

The remarkable feature of AD is that computing the gradient of a
scalar-valued function takes no more than five times as long as computing the
log posterior, \emph{regardless of the number of parameters}
\citep[p. xii]{GriewankWalther2008}.  If the cost of the log posterior
grows linearly in $N$, so will the cost of the gradient.

In most statistical software packages, like \proglang{R}, the default storage
mode for any matrix is in a ``dense'' format; each element in the
matrix is stored explicitly, regardless of the value.  For a model
with $n$ variables, this matrix consists of $n^2$ numbers, each
consuming eight bytes of memory at double precision.  If we have a
dataset in which $N=10000$, $k=5$ and $p$ is relatively small, the Hessian for
this model with $50,000+p$ variables will consume more than 20GB of RAM.  Furthermore, the
computational effort for matrix-vector multiplication is quadratic in
the number of columns, and matrix-matrix multiplication is cubic. To
the extent that either of these operations is used in the mode-finding
or sampling steps, the computational effort will grow much faster than
the size of the dataset.  Since multiplying a triangular matrix is
roughly one-sixth as expensive as multiplying a full matrix, we could gain some efficiency by
working with the Cholesky decomposition of the Hessian instead.  However, the
complexity of the Cholesky decomposition algorithm itself will still
be cubic in $N$ \citep[Ch. 1]{GolubVanLoan1996}.

For our purposes, the source of scalability is in the sparsity of the Hessian.
If the vast majority
of elements in a matrix are zero, we do not need to store them
explicitly.  Instead, we need to store only the non-zero values, the
row numbers of those values, and the index of the values that begin
each column.\footnote{This storage scheme is known as the Compressed
  Sparse Column (CSC) format.  This common format is used by the \pkg{Matrix}
  package in \proglang{R}, and the \pkg{Eigen} numerical library, but it is not the
  only way to store a sparse matrix.}  Under the
conditional independence assumption, the
cross-partial derivatives between heterogeneous parameters across
households are all zero.  Thus, the Hessian becomes sparser as the size of
the dataset increases.

\begin{figure}[tb]
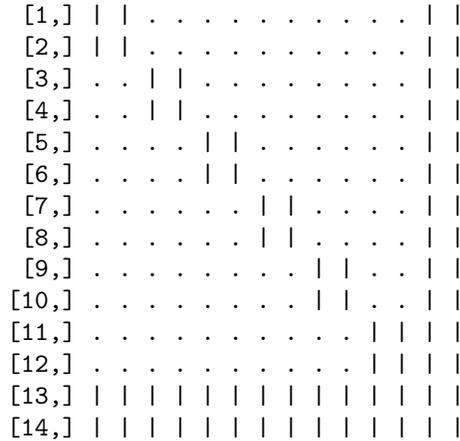
\small\centering
\begin{BVerbatim}
 [1,] | | . . . . . . . . . . | | 
 [2,] | | . . . . . . . . . . | |
 [3,] . . | | . . . . . . . . | |
 [4,] . . | | . . . . . . . . | |
 [5,] . . . . | | . . . . . . | | 
 [6,] . . . . | | . . . . . . | | 
 [7,] . . . . . . | | . . . . | | 
 [8,] . . . . . . | | . . . . | | 
 [9,] . . . . . . . . | | . . | | 
[10,] . . . . . . . . | | . . | |
[11,] . . . . . . . . . . | | | |
[12,] . . . . . . . . . . | | | |
[13,] | | | | | | | | | | | | | | 
[14,] | | | | | | | | | | | | | |
\end{BVerbatim}
 \caption{Example of sparsity pattern under conditional independence.}
\label{fig:blockArrow} 
\end{figure}

To illustrate, suppose we have a hierarchical model with six households, two heterogeneous parameters per
household, and two population-level parameters, for a total of 14
parameters.  Figure
\ref{fig:blockArrow} is a schematic of the sparsity structure of the Hessian;
the vertical lines are the non-zero elements, and the dots are the
zeros.  There are 196 elements in this matrix, but only 169 are
non-zero, and only 76 values are unique.  Although the savings in RAM
is modest in this illustration, the efficiencies are much greater when
we add more households.  If we had 1,000 households, with $k=3$ and
$p=9$, there would be 3009 parameters, and more than nine million
elements in the Hessian, yet no more than 63,000 are non-zero, of
which about 34,500 are unique.  As we add households, the number of
non-zero elements of the Hessian grows only linearly in $N$.

The cost of estimating a dense Hessian using AD grows linearly with
the number of variables \citep{GriewankWalther2008}.  When the Hessian is sparse,
with a pattern similar to Figure \ref{fig:blockArrow}, we can estimate
the Hessian so that the cost is only a multiple of the cost of computing the
log posterior.  We achieve this by using a graph coloring
algorithm to partition the variables into a small number of groups (or ``colors'' in
the graph theory literature), such that a small change in the variable
in one group does not affect the partial derivative of any other
variable in the same group.  This means we could perturb all of the
variables in the same group at the same time, recompute the gradient,
and after doing that for all groups, still be able to recover an
estimate of the Hessian.  Thus, the computational cost for computing
the Hessian grows with the number of groups, not the number of
parameters. Because the household-level parameters are conditionally
independent, we do not need to add groups as we add households.  For
the Hessian sparsity pattern in Figure \ref{fig:blockArrow}, we need
only four groups:  one for each of the heterogeneous parameters across
all of the households, and one for each of the two population-level
parameters.  In the upcoming binary choice example in Section \ref{sec:binaryChoice}, for which
$k=3$, there are
$\frac{1}{2}\left(k^2+5k\right)=12$ groups, no matter how many households
we have in the dataset.

\citet{CurtisPowellReid1974} introduce the idea of reducing
the number of evaluations to estimate sparse Jacobians.
\citet{PowellToint1979} describe how to partition variables into
appropriate groups, and how to recover Hessian information through
back-substitution.  \citet{ColemanMore1983} show that the task of
grouping the variables amounts to a classic graph-coloring
problem.  Most AD software applies this general principle to computing
sparse Hessians.  Alternatively, \proglang{R} users can use the \pkg{sparseHessianFD}
package \citep{R_sparseHessianFD} to efficiently estimate sparse
Hessians through finite differences of the gradient, as long as the
sparsity pattern is known in advance, and as long as the gradient was
not itself estimated through finite differencing.  This package is an
interface to the algorithms in \citet{ColemanGarbow1985} and \citet{ColemanGarbow1985b}.

\subsection{Finding the posterior mode}
For simple models and small datasets, standard default algorithms (like the \func{optim}
function in \proglang{R}) are sufficient for
finding posterior modes and estimating Hessians.  For larger problems,
one should choose optimization tools more thoughtfully.  For example,
many of the \proglang{R} optimization algorithms default to finite differencing of gradients when a gradient
function is not provided explicitly.  Even if the user can provide the
gradient, many algorithms will store a Hessian, or an approximation to
it, densely.  Neither feature is attractive when the number of
households is large.

For this section, let's assume that the log posterior is
twice-differentiable and unimodal.\footnote{Neither assumption is
  required, but most marketing models satisfy them, and
  maintaining them simplifies our exposition.} There are two approaches that one can take.  The first is to use a
``limited memory'' optimization algorithm that approximates the
curvature of the log posterior over successive iterations. Several
algorithms of this kind are described in \citet{NocedalWright2006},
and are available for many technical computing platforms.  Once the algorithm finds
the posterior mode, there remains the need to compute the Hessian exactly.

The second approach is to run a quasi-Newton algorithm, and compute the Hessian at each iteration
explicitly, but store the Hessian in a sparse format. The
\pkg{trustOptim} package for \proglang{R} \citep{R_trustOptim}
implements a trust region
algorithm that exploits the sparsity of the Hessian.  The user can
supply a Hessian that is derived analytically, computed using AD, or
estimated numerically using \pkg{sparseHessianFD}. Since memory
requirements and matrix computation costs will grow only linearly in $N$,
finding the posterior mode becomes feasible for large problems,
compared to similar algorithms that ignore sparsity.

We should note that we cannot predict the time to convergence for general
problems.  Log posteriors with ridges or plateaus, or that require
extensive computation themselves, may still take a long time to find
the local optimum.  Whether any mode-finding algorithm is ``fast
enough'' depends on the specific application.  However, if
one optimization algorithm has difficulty finding a mode, another
algorithm may do better.

\subsection{Sampling from an MVN distribution}
Once we find the posterior mode, and the Hessian at the mode,
generating proposal samples from an MVN($\theta^*,sH^{-1}$)  distribution is
straightforward.  Let $\frac{1}{s}H=\Lambda\Lambda'$ represent the
Cholesky decomposition of the precision of the proposal, and let $z$
be a vector of $Nk+p$ samples from a standard normal distribution.  To
sample $\theta$ from an MVN, solve the triangular linear system
$\Lambda'\theta=x$, and then add $\theta^*$.  Since $E(z)=0$,
$E(\theta)=\theta^*$, and since $E(zz')=I$,
$\cov(\theta)=\Lambda'^{-1}\Lambda^{-1}=(\Lambda\Lambda')^{-1}=sH^{-1}$.

Because $\Lambda$ is sparse, the costs of both
solving the triangular system, and the premultiplication, grow
linearly with the number of nonzero elements, which itself grows
linearly in $N$ \citep{Davis2006}.  If $\Lambda$ were dense, then the
cost of solving the triangular system would grow
quadratically in $N$.  Furthermore, computing the MVN density would involve
premultiplying $z$ by a triangular matrix, whose cost is cubic in $N$
\citep{GolubVanLoan1996}. 

Computation of the Cholesky decomposition can also benefit from the
sparsity of the Hessian.  If $H$ were dense $Nk+p$ square, symmetric matrix, then,
holding $k$ and $p$ constant, the
complexity order of the Cholesky decomposition is $N^3$
\citep{GolubVanLoan1996}. There are a number of different
algorithms that one can use for decomposing sparse Hessians \citep{Davis2006}.  The typical strategy is to first
permute the rows and
columns of $H$ to minimize the number of nonzero elements in
$\Lambda$, and then compute the sparsity pattern.  This part can be
done just once.  With the sparsity pattern in hand, the next step is
to compute those nonzero elements in $\Lambda$.  The time for this step grows with
the sum of the squares of the number of nonzero elements in each
column of $\Lambda$ \citep{Davis2006}.  Because each additional household adds
$k$ columns to $\Lambda$, with an average of $p+\frac{1}{2}(k+1)$ nonzero elements per
column, we can compute the sparse Cholesky decomposition in time that
is linear in $N$.  Software for sparse Cholesky
decompositions is widely available.

\subsection{Scalability Test}\label{sec:binaryChoice}

Next, we provide some empirical evidence of scalability
through a simulation study.  For a hypothetical dataset with $N$ households, let $y_i$ be
the number of times household $i$ visits a store during a $T$ week period. The
probability of a visit in a single week is $p_i$,
where $\logit p_i=\beta_i'x_i$, and $x_i$ is a vector of $k$ covariates.  The distribution of $\beta_i$ across
the population is MVN with mean $\bar{\beta}$ and covariance
$\Sigma$.  In all conditions of the test, we set $k=3$ and $T=52$, and
vary the number of households by setting $N$ to one of 10 discrete
values from 500 to 50,000.  The ``true'' values of
$\bar{\beta}$ are -10, 0 and 10, and the ``true'' $\Sigma$ is $0.1I$.
We place weakly informative priors on both $\bar{\beta}$ and
$\Sigma$.  

\begin{figure}[tb]\centering
    \begin{subfigure}[t]{.5\linewidth}\centering
     \includegraphics[scale=1,trim=0 10 0 10,clip]{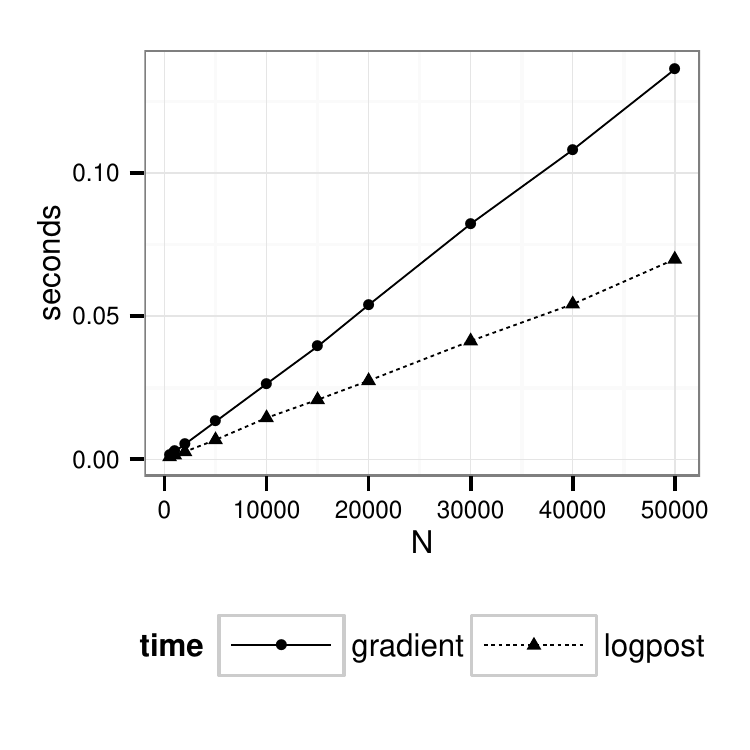}
    \end{subfigure}%
    \begin{subfigure}[t]{.5\linewidth}\centering
    \includegraphics[scale=1,trim=0 10 0 10,clip]{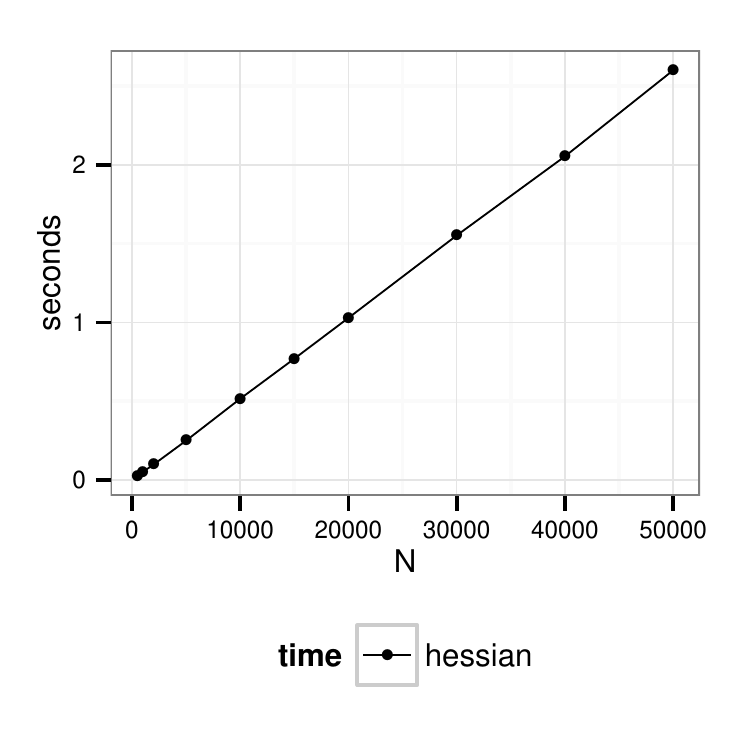}
  \end{subfigure}
  \caption{Average computation time for 100 evaluations of log posterior, gradient and Hessian.}
  \label{fig:logpost_time}
  \end{figure}

\begin{figure}[tb]\centering
   \includegraphics[scale=.8,trim=0 10 0 10,clip]{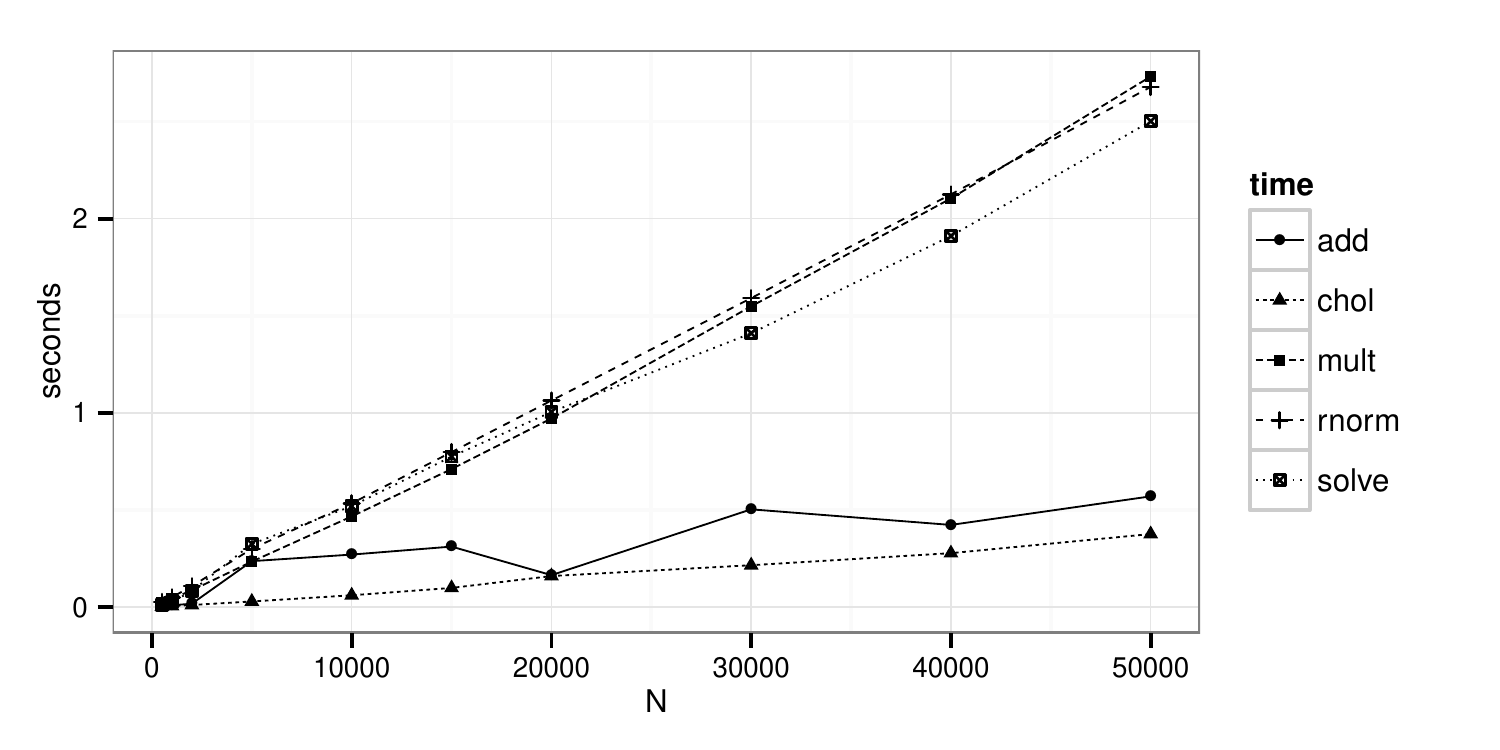}
   \caption{Computation time, averaged over 100 replications, for
     adding a vector to matrix columns (add);  a sparse Cholesky
     decomposition (chol); multiplying a sparse triangular matrix by a
     dense matrix (mult); sampling standard normal random variates (rnorm); and
     solving a sparse triangular linear system (solve)}
   \label{fig:mvn_time}
 \end{figure}

In Figure \ref{fig:logpost_time}, we plot the average time, across 100 replications, to
compute the log posterior, the gradient, and the Hessian.  As
expected, each of these computations grows linearly in $N$.  In Figure
\ref{fig:mvn_time}, we plot average times for the steps involved in sampling from an
MVN:  adding a vector to columns of a dense matrix, computing a sparse Cholesky decomposition,
multiplying a sparse triangular matrix by a dense matrix, sampling
standard normal random variates, and solving a sparse triangular
linear system. Again, we see that the time for all of these steps is
linear in $N$. 

Table \ref{tab:binaryAccRate} summarizes the acceptance rates and
scale factors when generating 50 samples from the posterior 
for different values of $N$.  Although there is a weak trend of increasing
acceptance rates with $N$,we cannot say with any certainty that
acceptance rates will always be either larger or smaller for larger
datasets.  The acceptance rate could be
influenced by using different scale factors on the Hessian for the MVN
proposal density.  However, we expect higher acceptance rates as
the target posterior density approaches an MVN asymptotically.  Since none of the steps in the algorithm grows
faster than linearly in $N$, we are confident in the scalability of
the overall algorithm.\begin{table}[tb]\small
\centering
\begin{tabular}{lcccccccccc}
N&500& 1,000 & 2,000 & 5,000 & 10,000 & 15,000 & 20,000 & 30,000&40,000& 50,000 \\
  \hline
scale factor&1.22& 1.16 & 1.10 & 1.08 & 1.04 & 1.03 & 1.03 &1.03&1.03& 1.02 \\ 
acc. rate ($\times 10^{-5}$) &2.1& 1.5& 2.6 & 0.7 & 2.9 & 2.6 &3.6&3.9& 2.7&
3.3\\
   \hline
\end{tabular}
\caption{Acceptance rates for binary choice simulation.  For each
  condition, $k=3$, so there are six population-level parameters and
  $3N$ heterogeneous parameters.}\label{tab:binaryAccRate}
\end{table}

\section{Estimating Marginal Likelihoods}\label{sec:MargLL}
Now, we turn to another advantage of our method: the
ability to generate accurate estimates of the marginal
likelihood of the data with little additional computation.  A number of researchers have proposed methods to
approximate the marginal likelihood, $\Ly$ from MCMC output
\citep{GelfandDey1994,NewtonRaftery94,Chib1995,RafteryNewton2007}, but
none has achieved universal acceptance as being consistent, stable and easy
to compute.  In fact, \citet{Lenk2009} demonstrated that methods that 
depend solely on samples from the posterior density could suffer from a
``pseudo-bias,'' and he proposes an importance-sampling method to correct for it.  This
pseudo-bias arises because the convex hull of MCMC samples defines only a
subset of the \emph{posterior} support, while $\Ly$ is defined as an
integral of the data likelihood over the \emph{prior} distribution. \citeauthor{Lenk2009} demonstrates that
his method dominates other popular methods, although with substantial
computational effort.  Thus, the estimation of the marginal likelihood remains a 
difficult problem in MCMC-based Bayesian statistics.

We estimate the marginal likelihood
using quantities that we already collect during the course of the
estimation procedure. Recall that $q(u)$ is the probability that, given a threshold value
$u$, a proposal from $\Gtheta$ is accepted as a sample from $\Post$.
Therefore, after substituting in Equation \ref{eq:margU}, we can express
the expected marginal acceptance probability for any one posterior sample as
\begin{align}
  \label{eq:5}
  \gamma&=\int_0^1q(u)p(u|y)~du =\frac{c_1}{c_2\Ly}\int_0^1q^2(u)~du
\end{align}
Applying a change of variables so $v=-\log u$, and then rearranging terms,
\begin{align}
  \label{eq:Ly}
  \Ly&=-\frac{c_1}{c_2\gamma}\int_0^{\infty}q^2(v)\exp(-v)~dv.
\end{align}

The values for $c_1$ and $c_2$ are immediately available from the
algorithm.  A reasonable estimator of $\gamma$ is $\hat{\gamma}$, the
inverse of the mean of the observed average number
of proposals per accepted sample.  What remains is estimating the integral in Equation \ref{eq:Ly}, for
which we use the same proposal draws that we already collected for estimating
$\widehat{q}_v(v)$.  The
empirical CDF of these draws is discrete, so we can partition the support of $q_v(v)$
at $v_1\mathellipsis v_M$.  Also, since $\widehat{q}_v(v)$ is the
proportion of proposal draws less than $v$, we have
$q_v(v_i)=i/M$. Therefore, 
\begin{align}
  \label{eq:9} \int_0^{\infty}q^2(v)\exp(-v)dv&\approx\sum_{i=1}^M\int_{v_i}^{v_{i+1}}\left(\frac{i}{M}\right)^2\exp(-v_i)dv\\
&=\frac{1}{M^2}\sum_{i=1}^M i^2\left[\exp(-v_i)-\exp(-v_{i+1})\right]\\
&=\frac{1}{M^2}\sum_{i=1}^M\left( 2i-1\right) \exp(-v_i)
\end{align}
Putting all of this together, we can estimate the marginal likelihood as
\begin{align}
  \label{eq:LyEst}
\Ly&\approx\frac{c_1}{M^2c_2\hat{\gamma}}\sum_{i=1}^M(2i-1)\exp(-v_i)
\end{align}

As a demonstration of the accuracy of this estimator, we use the
same linear regression example that \citet{Lenk2009} uses.
\begin{align}
  \label{eq:5}
  y_{it}&\sim N(x_i'\beta, \sigma^2),~i=1\mathellipsis n,
  t=1\mathellipsis T\\
\beta|\sigma&\sim N(\beta_0, \sigma^2V_0)\qquad
\sigma^2\sim IG(r,\alpha)
\end{align}
For this model, $\Ly$ is a multivariate-t density (MVT), which we can
compute analytically.  This allows us to compare the estimates of $\Ly$
with the ``truth.''  To do this, we conducted a simulation study for
simulated datasets of different numbers of observations $n\in\{200,2000\}$ and numbers of
covariates $k\in\{5, 25,100\}$.  For each $n,k$ pair, we simulated 25
datasets.  For each dataset, each vector $x_i$ includes an intercept and $k$ iid
samples from a standard normal density.  Thus, there are $k+2$
parameters, corresponding to the elements of $\beta$, plus $\sigma$.  The true intercept term is 5, and the remaining true
$\beta$ parameters are linearly spaced from $-5$ to $5$. In all cases,
there are $T=25$ observations per unit.  Hyperpriors are
set as $r=2$, $\alpha=1$, $\beta_0$ as a zero vector and $V_0=0.2\cdot
I_k$.  

For each dataset, we collected 250 samples from the posterior density,
with different numbers of proposal draws ($M = 1,000\text{ or }10,000$), and different scale factors ($s= 0.5, 0.6,
0.7,\text{ or } 0.8$) on the Hessian ($-sH$ is the precision matrix of the MVN
proposal density, and lower scale factors generate more diffuse
proposals).  We excluded the $s=0.8$, $n=200$ case because the proposal
density was not sufficiently diffuse to ensure that $\Phitheta$ was between 0 and
1 across the $M$ proposal draws.

Table \ref{tab:MargLL1} presents the true log marginal likelihood (MVT), along
with estimates using our method, the importance sampling method in Lenk
(2009), and the harmonic mean estimator \citep{NewtonRaftery94}.  We
also included the mean acceptance probabilities, and the standard
deviations of the various estimates across the simulated datasets. We can see that our estimates for the log
marginal likelihood are remarkably close to the MVT densities, and are 
robust when we use different scale factors. Accuracy
appears to be better for larger datasets than smaller ones, while improving the approximation of $p(u|y)$ by increasing
the number of proposal draws offers negligible improvement.  Note 
that the performance of our method is comparable to that of
Lenk, but is much better than the harmonic mean estimator. 
Our method is similar to Lenk's in that it computes the probability
that a proposal draw falls within the support of the posterior
density.  However, the inputs to the estimator of the marginal likelihood are
intrinsically generated as the algorithm progresses.  In contrast,
the Lenk estimator requires an additional importance sampling run
after the MCMC draws are collected.

\begin{table}[tbp]\small\centering
\begin{center}
\begin{tabular}{|rrrr|rrrrrrrr|r|}
  \hline
&&&&\multicolumn{2}{c}{MVT}&\multicolumn{2}{c}{Ours}&\multicolumn{2}{c}{Lenk}&\multicolumn{2}{c|}{HME}&Mean\\
k & n & M & scale & mean & sd & mean & sd & mean & sd &mean&sd&Acc \%\\
  \hline
 5 & 200 & 1000 & 0.5 & -309 & 6.6 & -309 & 6.6 & -311 & 6.8 & -287 & 7.1&22.1 \\ 
   5 & 200 & 1000 & 0.6 & -309 & 6.6 & -309 & 6.7 & -310 & 6.9 & -287 & 6.9&40.5 \\ 
   5 & 200 & 1000 & 0.7 & -309 & 6.6 & -309 & 6.7 & -310 & 6.5 & -287 & 6.3&57.1 \\ 
    \hline
 5 & 200 & 10000 & 0.5 & -309 & 6.6 & -309 & 6.6 & -311 & 6.7 & -287 & 6.7 &24.0\\ 
   5 & 200 & 10000 & 0.6 & -309 & 6.6 & -309 & 6.6 & -310 & 7.5 & -287 & 6.8&40.9 \\ 
   5 & 200 & 10000 & 0.7 & -309 & 6.6 & -309 & 6.7 & -310 & 7.0 & -287 & 7.1&55.2 \\ 
    \hline
 5 & 2000 & 1000 & 0.5 & -2866 & 46.2 & -2865 & 46.3 & -2868 & 46.2 & -2836 & 46.2&22.1 \\ 
   5 & 2000 & 1000 & 0.6 & -2866 & 46.2 & -2866 & 46.2 & -2868 & 45.7 & -2836 & 45.5&37.8 \\ 
   5 & 2000 & 1000 & 0.7 & -2866 & 46.2 & -2866 & 46.3 & -2867 & 45.9 & -2836 & 45.9&49.6 \\ 
   5 & 2000 & 1000 & 0.8 & -2866 & 46.2 & -2866 & 46.2 & -2867 & 46.3 & -2835 & 46.3&64.6 \\ 
   \hline
 5 & 2000 & 10000 & 0.5 & -2866 & 46.2 & -2866 & 46.4 & -2867 & 46.7 & -2836 & 46.9&25.3 \\ 
   5 & 2000 & 10000 & 0.6 & -2866 & 46.2 & -2866 & 46.2 & -2867 & 45.8 & -2836 & 46.3&36.3 \\ 
   5 & 2000 & 10000 & 0.7 & -2866 & 46.2 & -2866 & 46.4 & -2867 & 46.0 & -2836 & 46.3&51.4 \\ 
   5 & 2000 & 10000 & 0.8 & -2866 & 46.2 & -2866 & 46.2 & -2867 & 46.5 & -2835 & 46.3&72.0 \\ 
   \hline
25 & 200 & 1000 & 0.5 & -387 & 8.1 & -385 & 8.2 & -391 & 7.6 & -292 & 8.5&2.8 \\ 
  25 & 200 & 1000 & 0.6 & -387 & 8.1 & -386 & 8.1 & -390 & 9.5 & -292 & 8.8&8.1 \\ 
  25 & 200 & 1000 & 0.7 & -387 & 8.1 & -386 & 8.3 & -390 & 8.0 & -292 & 8.8 &16.2\\ 
    \hline
25 & 200 & 10000 & 0.5 & -387 & 8.1 & -385 & 8.5 & -390 & 8.2 & -292 & 8.4&1.7 \\ 
  25 & 200 & 10000 & 0.6 & -387 & 8.1 & -385 & 8.2 & -390 & 8.9 & -292 & 8.8&6.2 \\ 
  25 & 200 & 10000 & 0.7 & -387 & 8.1 & -386 & 8.2 & -390 & 8.7 & -292 & 9.1&20.0 \\ 
    \hline
25 & 2000 & 1000 & 0.5 & -2990 & 28.7 & -2989 & 28.8 & -2994 & 28.3 & -2865 & 28.8&2.7 \\ 
  25 & 2000 & 1000 & 0.6 & -2990 & 28.7 & -2989 & 28.7 & -2993 & 28.4 & -2864 & 29.0&4.6 \\ 
  25 & 2000 & 1000 & 0.7 & -2990 & 28.7 & -2989 & 28.9 & -2991 & 30.0 & -2864 & 29.5 &15.4\\ 
  25 & 2000 & 1000 & 0.8 & -2990 & 28.7 & -2990 & 28.7 & -2992 & 29.6 & -2864 & 29.4&43.1 \\ 
   \hline
25 & 2000 & 10000 & 0.5 & -2990 & 28.7 & -2988 & 29.2 & -2992 & 28.5 & -2864 & 28.9&.8 \\ 
  25 & 2000 & 10000 & 0.6 & -2990 & 28.7 & -2989 & 29.1 & -2993 & 29.4 & -2864 & 28.9&3.7 \\ 
  25 & 2000 & 10000 & 0.7 & -2990 & 28.7 & -2990 & 29.0 & -2993 & 28.9 & -2864 & 28.9&17.1 \\ 
  25 & 2000 & 10000 & 0.8 & -2990 & 28.7 & -2990 & 28.6 & -2993 & 28.2 & -2865 & 28.2&43.3 \\ 
   \hline
100 & 200 & 1000 & 0.5 & -660 & 6.7 & -661 & 6.5 & -683 & 8.8 & -292 & 9.2&.3 \\ 
  100 & 200 & 1000 & 0.6 & -660 & 6.7 & -660 & 6.6 & -678 & 8.5 & -286 & 9.0&.3 \\ 
  100 & 200 & 1000 & 0.7 & -660 & 6.7 & -659 & 7.1 & -673 & 7.8 & -282 & 8.0&.4 \\ 
    \hline
100 & 200 & 10000 & 0.5 & -660 & 6.7 & -659 & 6.9 & -682 & 9.1 & -288 & 10.4&.1 \\ 
  100 & 200 & 10000 & 0.6 & -660 & 6.7 & -660 & 5.7 & -678 & 8.8 & -286 & 8.9&.1 \\ 
  100 & 200 & 10000 & 0.7 & -660 & 6.7 & -658 & 6.7 & -674 & 7.3 & -282 & 8.4&.1 \\ 
    \hline
100 & 2000 & 1000 & 0.5 & -3364 & 24.4 & -3364 & 24.8 & -3370 & 27.5 & -2871 & 27.1&.3 \\ 
  100 & 2000 & 1000 & 0.6 & -3364 & 24.4 & -3362 & 24.6 & -3369 & 24.3 & -2868 & 25.3&.6 \\ 
  100 & 2000 & 1000 & 0.7 & -3364 & 24.4 & -3361 & 23.9 & -3371 & 25.6 & -2870 & 25.4&1.1 \\ 
  100 & 2000 & 1000 & 0.8 & -3364 & 24.4 & -3362 & 23.9 & -3370 & 26.0 & -2868 & 26.1&3.2 \\ 
   \hline
100 & 2000 & 10000 & 0.5 & -3364 & 24.4 & -3362 & 24.0 & -3372 & 25.3 & -2870 & 25.2&.1 \\ 
  100 & 2000 & 10000 & 0.6 & -3364 & 24.4 & -3360 & 24.9 & -3368 & 25.3 & -2867 & 25.4&.1 \\ 
  100 & 2000 & 10000 & 0.7 & -3364 & 24.4 & -3360 & 24.6 & -3370 & 25.5 & -2869 & 25.5&.4 \\ 
  100 & 2000 & 10000 & 0.8 & -3364 & 24.4 & -3362 & 24.5 & -3367 & 24.3 & -2867 & 24.4&3.0 \\ 
   \hline
\end{tabular}
\caption{Results of simulation study for effectiveness of estimator
  for log marginal likelihood.}\label{tab:MargLL1}
\end{center}
\end{table}

\section{Discussion of practical considerations and limitations}\label{sec:practical}

To those who have spent long work hours dealing with MCMC convergence and
efficiency issues, the utility of an alternative algorithm is appealing.  Ours allows for sampling
from a posterior density in parallel, without having to worry about
whether an MCMC estimation chain has converged.  If heterogeneous
units (like households) are conditionally independent, then the
sparsity of the Hessian of the log posterior lets us construct a
sampling algorithm whose complexity grows only linearly in the number
of units.  This method makes Bayesian inference
more attractive to practitioners who might otherwise be put off by the
inefficiencies of MCMC.

This is not to say that our method is guaranteed to generate perfect samples
from the target posterior distribution. One area of potential concern is that the empirical distribution
$\widehat{q}_v(v)$ is only a discrete
approximation to $q_v(v)$.  That discretization could introduce
some error into the estimate of the posterior density.  This error can
be reduced by increasing $M$ (the number of proposal draws that we use
to compute $\widehat{q}_v(v)$~), at the expense of costlier computation
of $\widehat{q}_v(v)$ and, possibly, lower acceptance rates.  In our
experience, and consistent with Figure \ref{fig:cheese}, we have not found this to be a problem, but some
applications for which this may be an issue might exist.

Like many other methods that collect random samples from posterior
distributions, its efficiency depends in part on a prudent selection
of the proposal density $g(\theta)$.  For the examples in this paper,
we use a MVN density that is centered at the posterior
mode with a covariance matrix that is proportional to the inverse of
the Hessian at the mode.  One might then wonder if there is an optimal
way to determine just how ``scaled out'' the proposal covariance needs to
be.  At this time, we think that manual search is the best alternative.  If we start with a small $M$ (say,
100 draws), and find that $\Phitheta>1$ for any of the $M$ proposals,
we have learned that the proposal density is not valid, with little
computational or real-time cost.  We can then re-scale the proposal
until $\Phitheta<1$, and then gradually increase $M$ until we get a
good approximation to $p(u)$.  This is no different, in principle,
than the tuning step in a Metropolis-Hasting algorithm.  However, our method has the advantage that we can make these
adjustments before
the posterior sampling phase begins. In contrast, 
with adaptive MCMC methods an improperly tuned sampler might not
be apparent until the chain has run for a substantial period of time.
Also, even if an acceptance rate appears to be low, we can still collect
draws in parallel, so the ``clock time'' remains much less than the
time we spend trying to optimize selection of the proposal. 

There are many popular models, such as multinomial probit, for which the
likelihood of the observed data is not available in closed form.  When
direct numerical approximations to these likelihoods (e.g., Monte
Carlo integration) are not tractable, MCMC with data
augmentation is a possible alternative.  Recent advances 
in parallelization using graphical processing units (GPUs) might make numerical estimation of integrals
more practical than it was even
10 years ago \citep{SuchardWang2010}.  If so, then our method could be a viable, efficient alternative to data
augmentation in these kinds of models.  Multiple
imputation of missing data could suffer from the same kinds of
problems, since a latent parameter, introduced
for the data augmentation step, is only weakly identified on its own.
If the number of missing data points is small, one could treat the
representation of the missing data points as if they were parameters.
But the implications of this require additional research.

Another opportunity for further research involves the case of multimodal posteriors.  Our method
does require finding the global posterior mode, and all the models
discussed in this paper have unimodal posterior distributions. When
the posterior is multimodal, one might instead use a mixture of
normals as the proposal distribution.
The idea is to not only find the global mode, but any local ones as
well, and center each mixture component at each of those local
modes. The algorithm itself would remain unchanged, as long as the global
posterior mode matches the global proposal mode.

We recognize that finding all of the local modes could be a hard
problem, and there is no
guarantee that any optimization algorithm will find all local
extrema in a reasonable amount of time.  In practical terms, MCMC,
offers no such guarantees either.  Even if the log posterior density
is unimodal, one should take care that the
mode-finding optimizer does not stop until reaching the optimum.  For
\proglang{R}, \pkg{trustOptim} \citep{R_trustOptim} is one such
package, in that its stopping rule depends on the norm of the gradient
being sufficiently close to zero. 

There are a number
of packages  for the \proglang{R} statistical programming language that can help with implementation of our method.  The \pkg{bayesGDS}
package \citep{R_bayesGDS} includes functions to run the rejection sampling phase (lines
20-36 in Algorithm \ref{alg:GDS}).  This package
also includes a function that estimates the log marginal likelihood
from the output of the algorithm.  If the
proposal distribution is MVN, and either the covariance or precision
matrix is sparse, then one can use the \pkg{sparseMVN} \citep{R_sparseMVN} package to sample
from the MVN by taking advantage of that sparsity.  The
\pkg{sparseHessianFD} \citep{R_sparseHessianFD} package estimates a sparse Hessian 
by taking finite differences of gradients of the function, as long as
the user can supply the sparsity pattern (which should be the case
under conditional independence).  Finally, the \pkg{trustOptim}
package \citep{R_trustOptim} is a nonlinear optimization package that uses a sparse Hessian to include
curvature information in the algorithm.

\singlespacing
\linespread{1.05}

\section*{Acknowledgment}
The authors acknowledge research assistance from Jonathan Smith, and are grateful for helpful suggestions and comments from Eric
Bradlow, Peter Fader, Fred Feinberg,  John Liechty,  Blake McShane,
Steven Novick,  John Peterson, Marc Suchard, Stephen Walker and Daniel Zantedeschi.

\printbibliography


\end{document}